# Perspective: Highly stable vapor-deposited glasses

M.D. Ediger, Department of Chemistry, University of Wisconsin-Madison, 1101 University Avenue, Madison, WI 53706 USA

**Abstract:** This article describes recent progress in understanding highly stable glasses prepared by physical vapor deposition and provides perspective on further research directions for the field. For a given molecule, vapor-deposited glasses can have higher density and lower enthalpy than any glass that can be prepared by the more traditional route of cooling a liquid, and such glasses also exhibit greatly enhanced kinetic stability. Because vapor-deposited glasses can approach the bottom of the amorphous part of the potential energy landscape, they provide insights into the properties expected for the "ideal glass". Connections between vapor-deposited glasses, liquid-cooled glasses, and deeply supercooled liquids are explored. The generality of stable glass formation for organic molecules is discussed along with the prospects for stable glasses of other types of materials.

**Introduction.** Glasses are nonequilibrium amorphous solids that play a critical role in modern technology. While everyone is familiar with window glass (a silicate-based oxide glass), not everyone will identify that many polymeric solids are glasses (such as plexiglass and polystyrene). Metallic glasses are utilized in the core of electrical transformers because they more effectively minimize losses in comparison to polycrystalline metals. Most smart phone displays are OLEDs (organic light emitting diodes), with more than 300 million such displays manufactured each year. The active elements in OLEDs are vapor-deposited glasses of organic semiconductors;[1] these layers transport holes and electrons, and allow these charges to recombine on an organic emitter to produce light.

It is worth reflecting on why a glass would be used for some applications rather than a crystal. In spite of their local disorder, glasses can be spectacularly homogenous on a macroscopic scale (e.g., on the length scale relevant for transmission of visible light). Optical fibers are made of glass and form the backbone of our modern communication network. These fibers can transmit optical signals more than 100 km without amplification; this could never be accomplished with a crystalline material because of scattering from grain boundaries. Glasses are flexible about composition in a way that crystals generally are not. While crystals are typically very sensitive to impurities and often exclude them from the crystal lattice, glass formulations often contain a significant fraction of several components without phase separation. In addition, because glasses are out of equilibrium, the process of formation can be manipulated to produce the glass that best suits the application. OLED displays make use of the three properties of glass highlighted in this paragraph. Homogeneity ensures that each pixel of a particular color emits the same amount of light. The emitter layer is typically a mixture of at least two molecules. And the substrate temperature during vapor deposition can be selected to produce glasses in which the emitter molecules have transition dipoles in the plane of the device, thus maximizing emission efficiency.

The goal of this article is to describe recent progress in understanding the highly stable glasses that can be prepared by physical vapor deposition. While vapor deposition has been used to prepare organic glasses for several decades,[2, 3] only in the last 10 years has it been realized that some deposition conditions produce truly remarkable materials.[4] For a given molecule, vapor-deposited glasses can be higher in density and modulus than any glass that can be prepared by the more traditional route of cooling a liquid. Such glasses are near the limits of what is possible for amorphous packing arrangements. In addition, vapor-deposited glasses can exhibit greatly enhanced kinetic stability and are often described as "stable glasses". The properties of liquid-cooled glasses evolve with time in a manner that is limiting for some applications and so improving the stability of glasses is practically important. While vapor deposition is an important industrial route to glasses (for cell phone displays and OLED televisions) and will likely play a role in additional applications of organic electronics, the emphasis of this article will be more fundamental: *What can vapor-deposited glasses teach us about supercooled liquids and the extreme properties of amorphous solids?*

**Extreme amorphous materials.** A good starting point for understanding vapor deposition is to describe the most common method of fabricating glass – the cooling of a liquid. Figure 1 shows the molar volume as a function of temperature for a particularly good glassformer (tris-naphthylbenzene)[5] but the qualitative features shown are generic for organic glassformers. Equilibrium states are shown in black; these lines illustrate the volume decrease that occurs upon crystallization during cooling. If the liquid does not crystallize upon cooling, the supercooled liquid is formed. The supercooled liquid is stable as long as no crystal nuclei are formed; the supercooled liquid is a metastable state and is shown in blue. In the temperature range shown by the solid blue line, the molecular reorientation time increases from ~1 ns to ~1000 s. Upon cooling at a constant rate (1 K/min), a temperature is reached where molecular motions are so slow that the system can no longer stay in equilibrium and a glass is formed; this is the glass transition temperature $T_g$. The molar volumes of two particular glasses are shown in red, with the color designating that these are non-equilibrium states. The upper red curve shows a glass prepared by cooling at 1 K/min while the lower curve shows a denser glass prepared by annealing the first glass for 4 days at $T_g$ – 10 K. This change in density over 4 days is called aging (or “physical aging” to make clear that it is not a result of chemical changes) and illustrates that under these conditions the liquid-cooled glass is barely stable.

There is an important region in Figure 1 which is inaccessible for liquid-cooled glasses (marked with “?”). These high density glasses cannot practically be reached by isothermal aging because aging slows logarithmically in time. Whatever densification occurs via aging between t = 1 hour and t = 100 hours (i.e., the difference between the two red lines in Figure 1), one would have to wait until t = 10000 hours to see another increase in density by this amount.[6] As we discuss below, physical vapor deposition allows access to this unexplored region of Figure 1, with the preparation of such materials requiring less than 1 hour. At present, this is the only way to produce these dense glasses on human time scales.

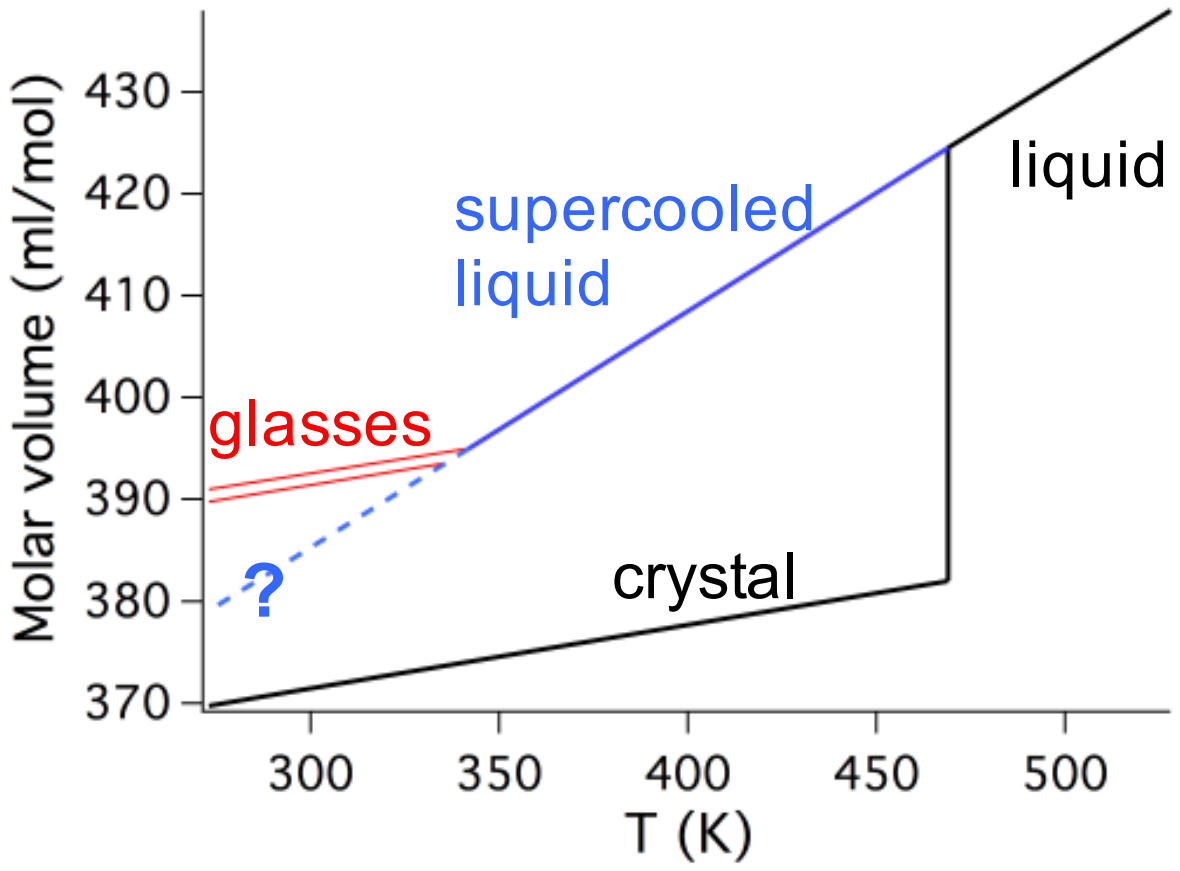

*Figure 1. Molar volume of tris-naphthylbenzene (1,3-bis(1-naphthyl)-5-(2-naphthyl)benzene) in the crystal, liquid and supercooled liquid states. In addition, the molar volumes of two glasses are also shown. The data in the figure comes from reference [5].*

Why would one care about the "?" region of Figure 1? Glasses in this regime are expected to have extreme properties. As a liquid-cooled glass is aged, not only does the density increase, but also the modulus and thermal stability increase, and aging makes glasses less permeable to environmental gases. The glasses in the "?" regime will have extreme properties associated with the best possible amorphous packing.

While Figure 1 gives the impression that the density of a supercooled liquid might be increased without bound if one could cool slowly enough to stay in equilibrium down to very low temperatures, this is probably not the case. Kauzmann pointed out in 1948 that there is a limit imposed on this process by the entropy,[7] as illustrated in Figure 2. Here the molar entropy of o-terphenyl is presented in the crystal, the liquid, and the supercooled liquid, as obtained from calorimetry measurements.[8] The blue dashed line is a naive extrapolation of the entropy of the supercooled liquid, i.e., we would expect to stay on this line down to low temperature if only we could cool slowly enough to remain in equilibrium. As discussed by Kauzmann, this extrapolation is problematic. At about 70 K, the absolute entropy of o-terphenyl would become negative, in violation of the laws of thermodynamics. Even at much higher temperatures, the entropy of the extrapolated supercooled liquid equals that of the crystal (at a temperature denoted as $T_K$). It is difficult to imagine an amorphous packing with entropy as low as that of the crystal, particularly as the supercooled liquid is expected to have a lower density. Workers in the field regard $T_K$ as a real limit for the supercooled liquid, in the following sense: If, as the temperature is lowered, the entropy of the supercooled liquid continues to follow the extrapolation from higher temperature, then that extrapolation must fail near $T_K$ or at a higher temperature.[9] One possibility is a second order phase transition to an "ideal glass" state; in this case, the entropy would follow the dashed extrapolation down to $T_K$ and then become equal to the crystal entropy at lower temperatures.[10] In this scenario, there is a well-defined limit to amorphous packing and in order to achieve it, the supercooled liquid would need to stay in equilibrium down to $T_K$ (about 85% of the conventional $T_g$, for o-terphenyl).

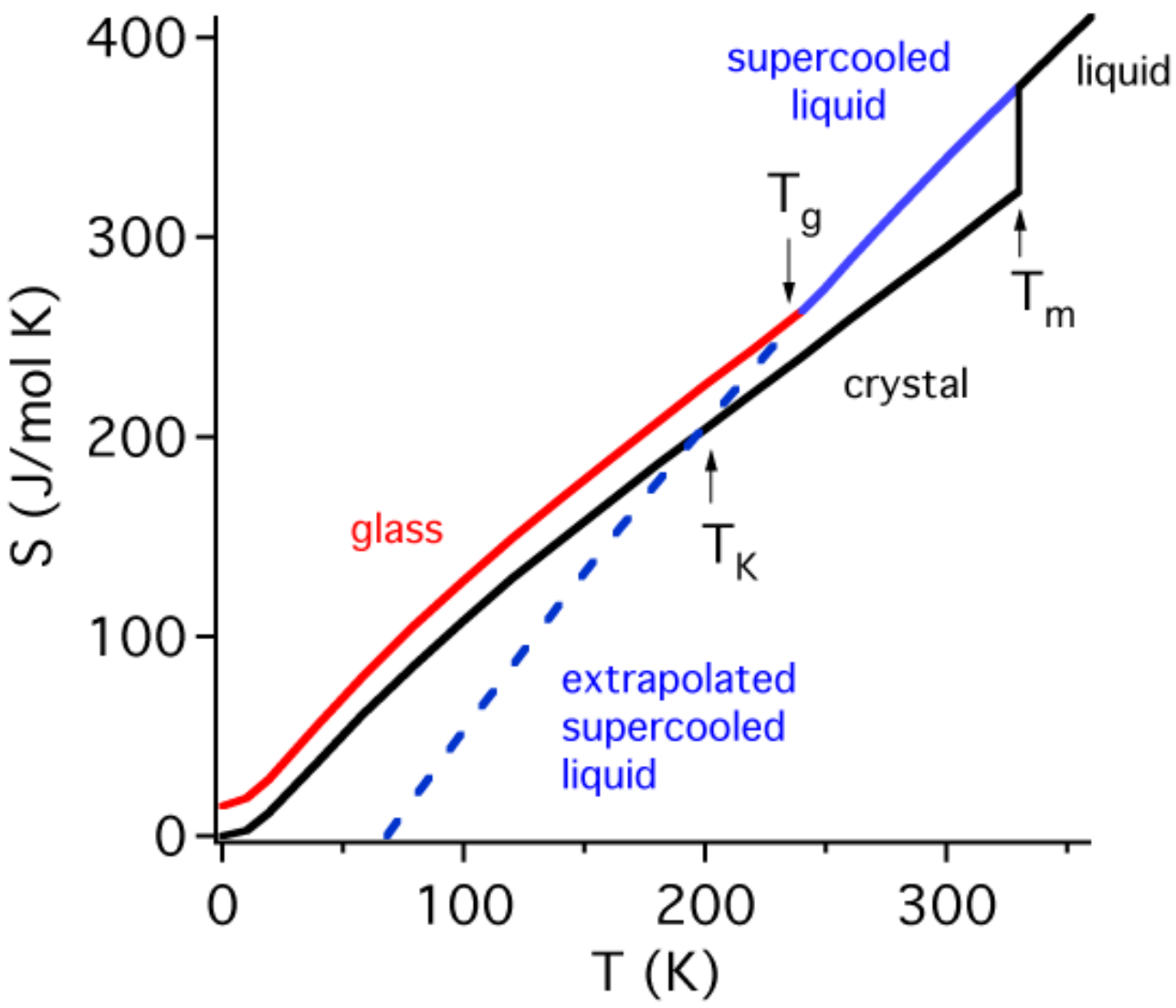

*Figure 2. Molar entropy of o-terphenyl (OTP) in the crystal, liquid, and supercooled liquid states; also shown is the estimated entropy of a liquid-cooled glass.*[8] *The dashed blue line is an extrapolation of the supercooled liquid entropy to lower temperature. Reproduced from J. Chem. Phys.* ***137****, 080901 (2012), with the permission of AIP Publishing.*

The potential energy landscape controls the dynamics, thermodynamics, and structure of amorphous systems – and provides an important way to think about the properties of optimally-packed amorphous solids.[11, 12] While this landscape is a function of the coordinates of all the molecules in the system, we can illustrate some of its important features in a one-dimensional representation as shown in Figure 3.[12] Basins represent possible states of the system with potential energy barriers acting as the transition states. As the temperature of the supercooled liquid is lowered, the average potential energy lowers. "$T_A$" marks the temperature where non-Arrhenius dynamics are first observed; here the characteristic time scale for molecular reorientation is typically on the order of nanoseconds. "$T_g$" marks the level reached on the landscape when cooling the liquid at 1 K/min; when the material leaves equilibrium at $T_g$, it is stuck on the energy landscape and further cooling does not lower its potential energy. Extended aging not too far below $T_g$ allows some equilibration and lowers the position of the system on the landscape; the line shown might indicate the influence of one year of aging. The position on the energy landscape achieved for some systems by physical vapor deposition is much lower than can be achieved in the laboratory by aging or slow cooling, as we discuss below.

A key feature of Figure 3 is that there is a well-defined bottom to the amorphous part of the potential energy landscape. The rapid drop of the entropy of the supercooled liquid as it is cooled indicates a landscape with fewer and fewer amorphous states at lower potential energy, i.e., fewer and fewer ways of arranging the molecules. The bottom of the landscape corresponds to the temperature where the part of the entropy associated with configurations ($S_{conf}$) reaches zero. This represents a limiting state of perfected amorphous packing which is sometimes called the "ideal glass". Whether the actual landscapes of real systems have an ideal glass state (with $S_{conf}$ =0) is an open question that we discuss below. While some models of the glass transition make specific predictions about what happens near the bottom of the landscape,[10, 13, 14] the properties of glasses near the bottom of the landscape are of interest beyond these models. For example, it is likely that the lowest energy glass has the highest modulus and the highest thermal stability of any amorphous packing. For cooling at constant pressure, the lowest energy glass is likely the densest possible amorphous packing of a given system. As discussed below, states approaching the ideal glass or lowest energy glass can be prepared by physical vapor deposition.

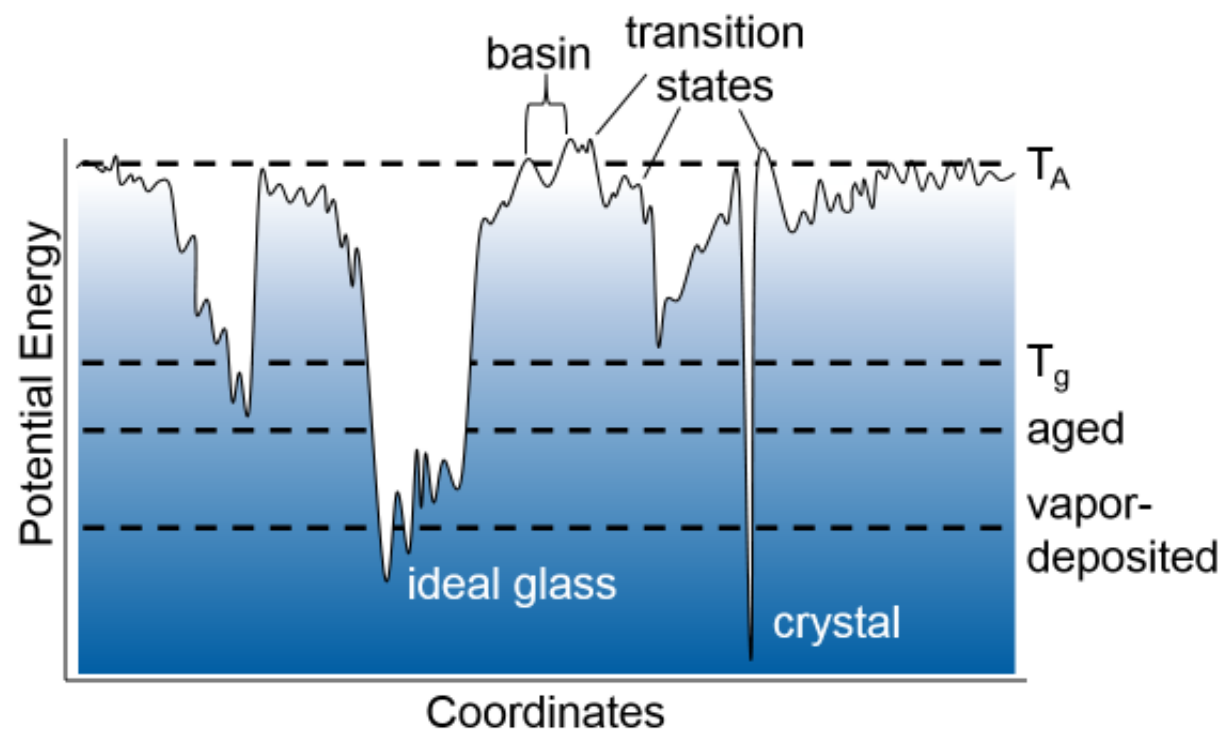

*Figure 3. A schematic representation of the potential energy landscape of a glassforming system. At constant volume, a single landscape controls the dynamics, thermodynamics, and structure of the supercooled liquid and all possible glassy states.*

**High density and low enthalpy vapor-deposited glasses.** Figure 1 highlights an important regime of high density materials that is inaccessible from the supercooled liquid because of the enormous times scales that would be required for cooling or aging. In 2007, Swallen et al. showed that glasses with these high densities could be prepared by physical vapor deposition.[4] Figure 4 shows results from an ellipsometry experiment on a ~600 nm film of indomethacin.[15] The glass was vapor-deposited onto a silicon substrate held at $T_{substrate}$ = 285 K and then heated at 1 K/min while measuring the thickness. After the sample transformed into the supercooled liquid, it was cooled and then re-heated at 1 K/min in order to obtain data on a reference liquid-cooled glass. The as-deposited glass is 0.85% thinner (i.e., 0.85% more dense) than the liquid-cooled glass. An extrapolation of the data shown indicates that the density of the as-deposited glass (at $T_{substrate}$) is very close to that expected for the equilibrium liquid at this temperature, even though this is 25 K below the conventional $T_g$ = 310 K.   There are a number of ways to estimate how long one would have to age a liquid-cooled glass to attain this density, with results ranging from 100 to 100,000 years.  Based upon these estimates, such high-density glasses until recently would have been considered “impossible materials”. Another notable feature of the as-deposited glass in Figure 4 is its high kinetic stability as indicated by the fact that the onset temperature for transformation into the supercooled liquid is 20 K higher than for the liquid-cooled glass; this indicates that the energy barriers governing rearrangement are higher in the as-deposited glass.  Qualitatively, one could consider the as-deposited glass to be “superaged” in that it has the high density and high kinetic stability expected for highly aged liquid-cooled glasses. As shown below, even denser and “older” glasses of indomethacin can be prepared at lower $T_{substrate}$.

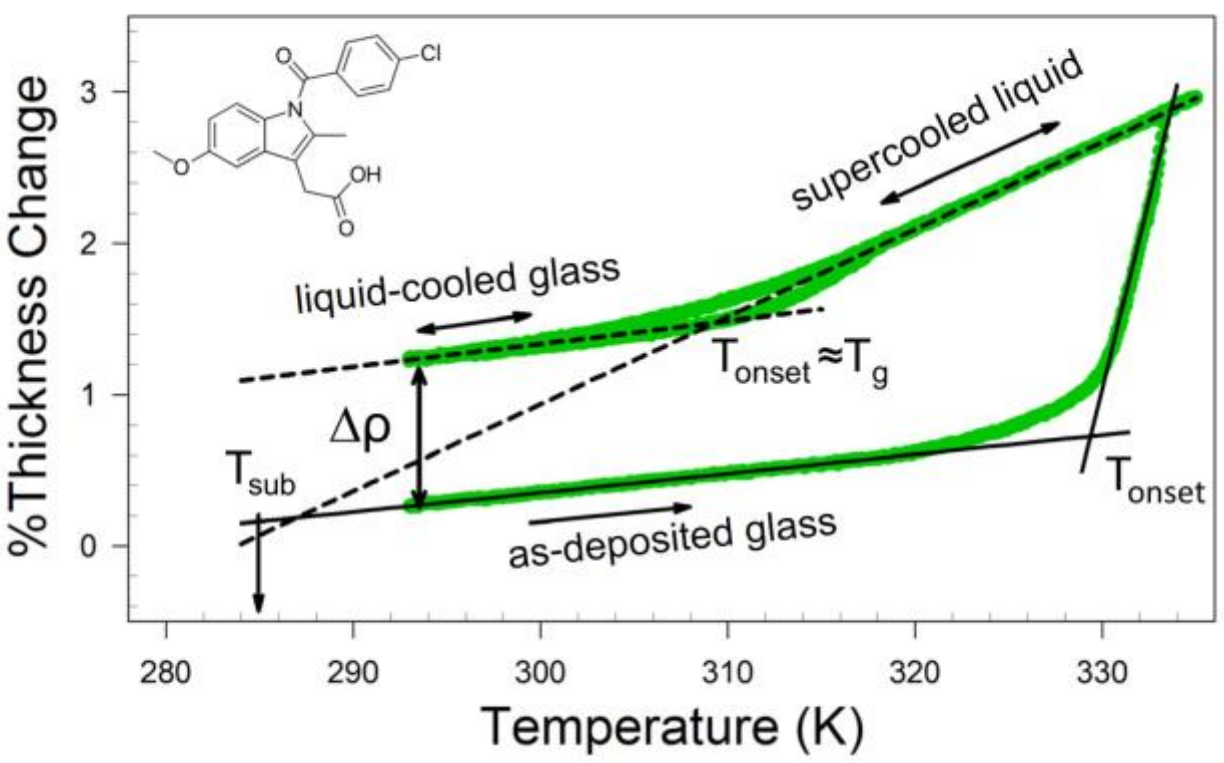

*Figure 4. Spectroscopic ellipsometry measurements of film thickness for a glass of indomethacin vapor-deposited at 0.2 nm/s onto a silicon substrate at 285 K. The arrows indicate the thermal path of the sample, with heating and cooling at 1 K/min. The as-deposited glass is denser and has a higher onset temperature than the liquid-cooled glass. Data from reference [15].*

The results shown in Figure 4 are surprising given the pre-2007 literature on vapor deposition. It had been thought that vapor-deposited glasses always exhibited low density and low stability, as a result of the very fast cooling rate for individual molecules when they hit the surface (~ $10^{13}$ K/s). From Figure 1, we see that fast cooling rates would be expected to yield glasses with low density since the system would leave equilibrium at a high temperature. The formation of high density glasses via vapor deposition can be rationalized by a surface equilibration mechanism: If there is sufficiently high mobility at the surface of the glass, molecules can equilibrate (or nearly equilibrate) as they are deposited even if the temperature is below $T_g$.[4, 16] Even in 2007, there were indications that glass surfaces might be more than $10^4$ times faster than the bulk.[17] More recent work on molecular glass formers has shown that surface diffusion can be up to $10^8$ times faster than bulk diffusion at $T_g$.[18-22] Thus, even though each molecule is only near the surface for ~ 10 s during the deposition, the dynamics are so fast that this provides sufficient configurational sampling to get close to equilibrium. Further deposition locks molecules into place, as once they are more than a few monolayers from the surface, they no longer benefit from high surface mobility. As every molecule has an opportunity to participate in efficient configurational sampling, vapor deposition can result in a tightly packed bulk material in which extremely high energy barriers must be surmounted to allow rearrangement.

Whether a high-density glass can be formed by physical vapor deposition depends upon the deposition rate and the substrate temperature during deposition. The effect of the substrate temperature is illustrated in Figure 5. Here indomethacin was vapor-deposited onto a substrate with a temperature-gradient imposed upon it.[23] In this manner, a library of glasses could be prepared in one deposition; this library explores the effect of substrate temperature at constant deposition rate (~ 0.2 nm/s). The vertical axis shows the density of the as-deposited glass relative to the density of a reference glass cooled from the liquid at 1 K/min (with both measured at room temperature). The features of this plot are easily understood given the surface equilibration mechanism discussed above. The red line shows the density expected for the equilibrium supercooled liquid (by extrapolation from above $T_g$). We interpret the data on the right side of the graph as a region of thermodynamic control; mobility is high enough to allow the equilibrium density to be attained. We interpret the data on the left side as a region of kinetic control; presumably mobility at the surface is so low that equilibration is strongly limited. The glasses with the maximum density were deposited at an intermediate substrate temperature; here there is enough mobility to nearly achieve the very large density expected at equilibrium. For deposition above $T_g$, we expect and observe very fast equilibration to the supercooled liquid during deposition, and then formation of a typical liquid-cooled glass upon cooling at the end of the deposition. The effect of deposition rate has been explored for density and other properties.[16, 24-26] As expected for the surface equilibration mechanism, lowering the deposition rate has no influence for substrate temperatures just below $T_g$,, while in the kinetically controlled region, slower deposition generally results in properties associated with more highly equilibrated glasses.[24, 27]

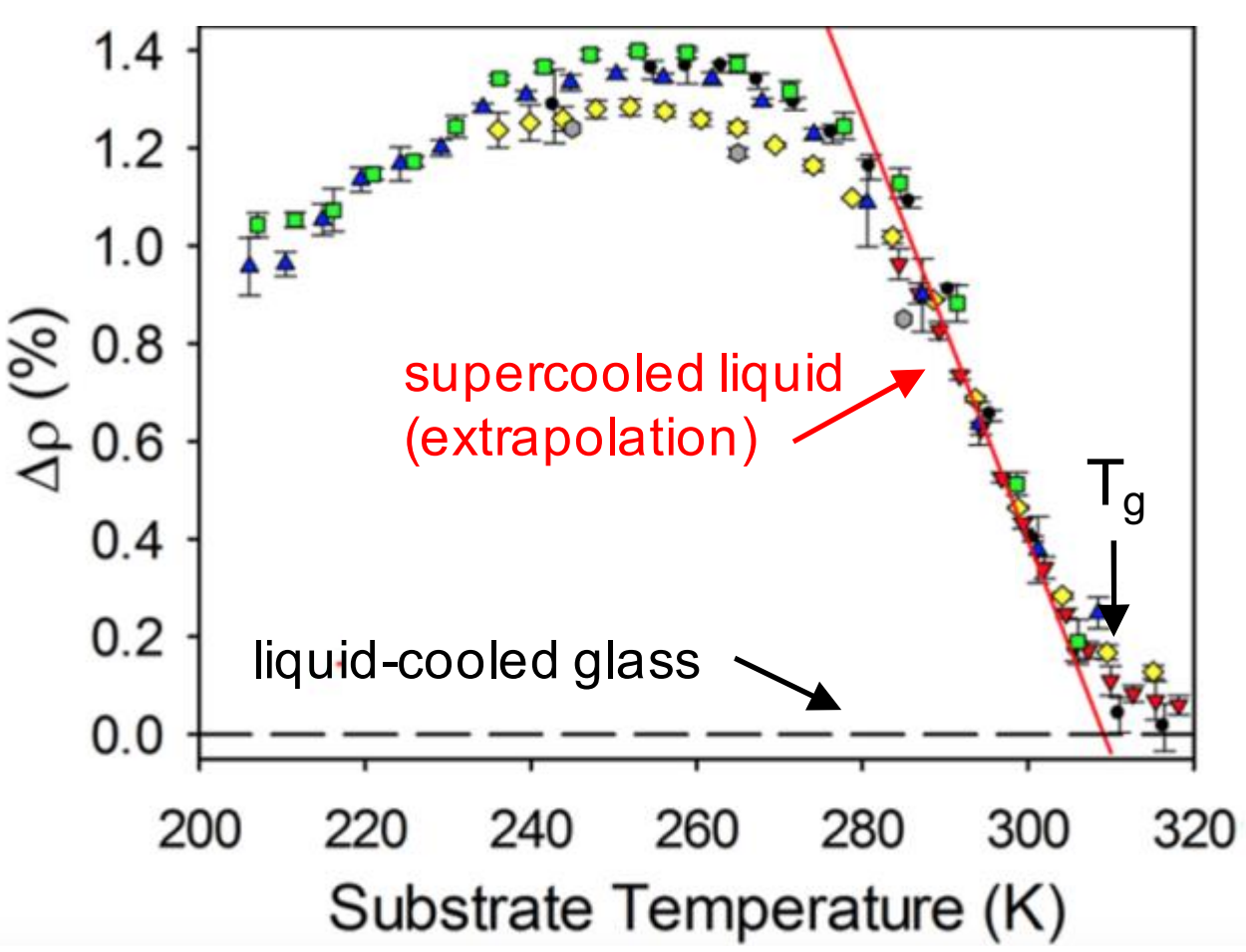

*Figure 5. Densities of vapor-deposited glasses of indomethacin relative to the liquid-cooled glass, as a function of the substrate temperature during deposition. The red line indicates the density expected for the equilibrium supercooled liquid (obtained by extrapolation). The five different colors correspond to five different temperature gradient samples. Reproduced with permission from J. Phys. Chem. B* ***117****, 15415-15425 (2013). Copyright 2013 American Chemical Society.*

The surface equilibration mechanism for stable glass formation has been tested in computer simulations of vapor deposition.[28-30] For a polydisperse mixture of Lennard-Jones particles, Berthier et al. compared the potential energies of glasses prepared by deposition at various rates with those produced by cooling the liquid. As expected given the above mechanism, their results indicated that “surface relaxation plays the same role in the formation of vapor-deposited glasses as bulk relaxation in ordinary glass formation”.[30] No similar test has been performed experimentally but this is an important goal given recent indications that the surface diffusion coefficient might not be the most relevant descriptor of surface mobility for the vapor deposition process.[31]

Ramos et al. performed high precision adiabatic calorimetry experiments on ethylbenzene that illustrate the efficiency with which vapor deposition can prepare glasses that are low in the energy landscape.[32] Figure 6 shows the enthalpy of six vapor-deposited glasses (open circles). The uppermost symbol is the enthalpy of a liquid that was cooled very slowly (10 mK/min), at the lowest temperature at which equilibrium could be maintained (112.5 K). The solid line is the expected enthalpy of the supercooled liquid based upon extrapolation from higher temperatures. The samples vapor-deposited with $T_{substrate}$ down to 105 K all have enthalpies consistent with that expected for the equilibrium supercooled liquid. Figures 5 and 6 tell the same story, for two different properties. Figure 6 can also be interpreted as indicating a regime of thermodynamic control for depositions at 105 K and above, with kinetics controlling the enthalpy of glasses deposited at lower temperature. Based upon the temperature dependence of the dielectric relaxation time,[33] one can estimate that, starting from a liquid-cooled glass, it would require between $10^4$ and $10^{13}$ years of aging or slow cooling to reach the enthalpy of the sample vapor-deposited at 105 K. Results similar to those shown in Figure 6 have been obtained for vapor-deposited toluene by Leon-Gutierrez et al;[34, 35] in this case, the vapor-deposited glasses match the enthalpy expected for the supercooled liquid down to 112 K (the liquid-cooled glass shows $T_g$ = 117 K).

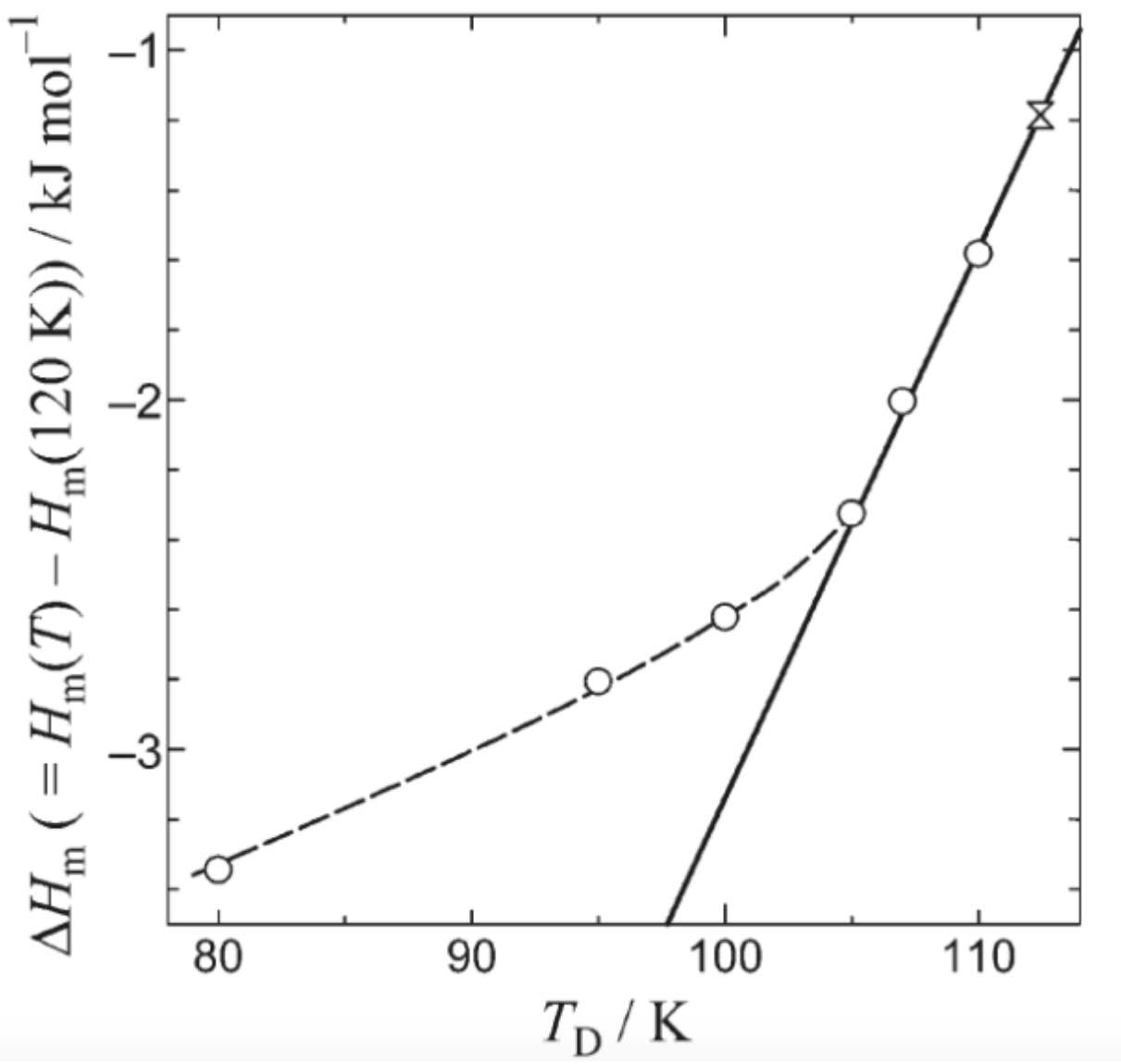

*Figure 6. Enthalpies of vapor-deposited glasses of ethylbenzene as a function of substrate temperature during deposition (o). The solid line is the expected enthalpy of the supercooled liquid obtained by extrapolation. The uppermost point is the enthalpy of the supercooled liquid at the lowest temperature for which equilibrium could be attained. Reproduced with permission from [32]. Reproduced with permission from J. Phys. Chem. B* ***115****, 14327-14332 (2011). Copyright 2011 American Chemical Society.*

**The entropy crisis revisited**. The data in Figures 5 and 6 provide an exciting opportunity to examine the entropy crisis in supercooled liquids at significantly lower temperatures than has been previously possible. The data in these figures are consistent with the idea that the properties of vapor-deposited glasses in the thermodynamic control regime are equal to those expected for the equilibrium supercooled liquid. Assuming that this is not a coincidence (and this should be tested further by experiments at lower deposition rates), this indicates that the extrapolated enthalpy of the supercooled liquid is correct in the thermodynamic control regime. If the extrapolated enthalpy is correct, then the extrapolated $C_p$ is correct, and the extrapolated S must also be correct in this regime. Figure 7 shows the configurational entropies of the supercooled liquids of ethylbenzene and toluene as a function of temperature. Here $S_{conf} = S - S_{vib}$ with $S_{vib}$ is estimated from heat capacity measurements on the liquid-cooled glass.[36] The open symbols show values of $S_{conf}$ obtained from adiabatic calorimetry by Yamamuro et al.[36] These measurements were performed on the supercooled liquid prepared by cooling from above $T_m$; the lowest temperature shown for the open symbols represents the lowest temperature at which the supercooled liquid remained in equilibrium. The asterisks indicate entropies calculated from vapor-deposited glasses[32, 34, 35] according to the argument described above. The line through both data sets is a fit allowing an extrapolation to $T_K$ (here defined as the temperature where $S_{conf} = 0$).

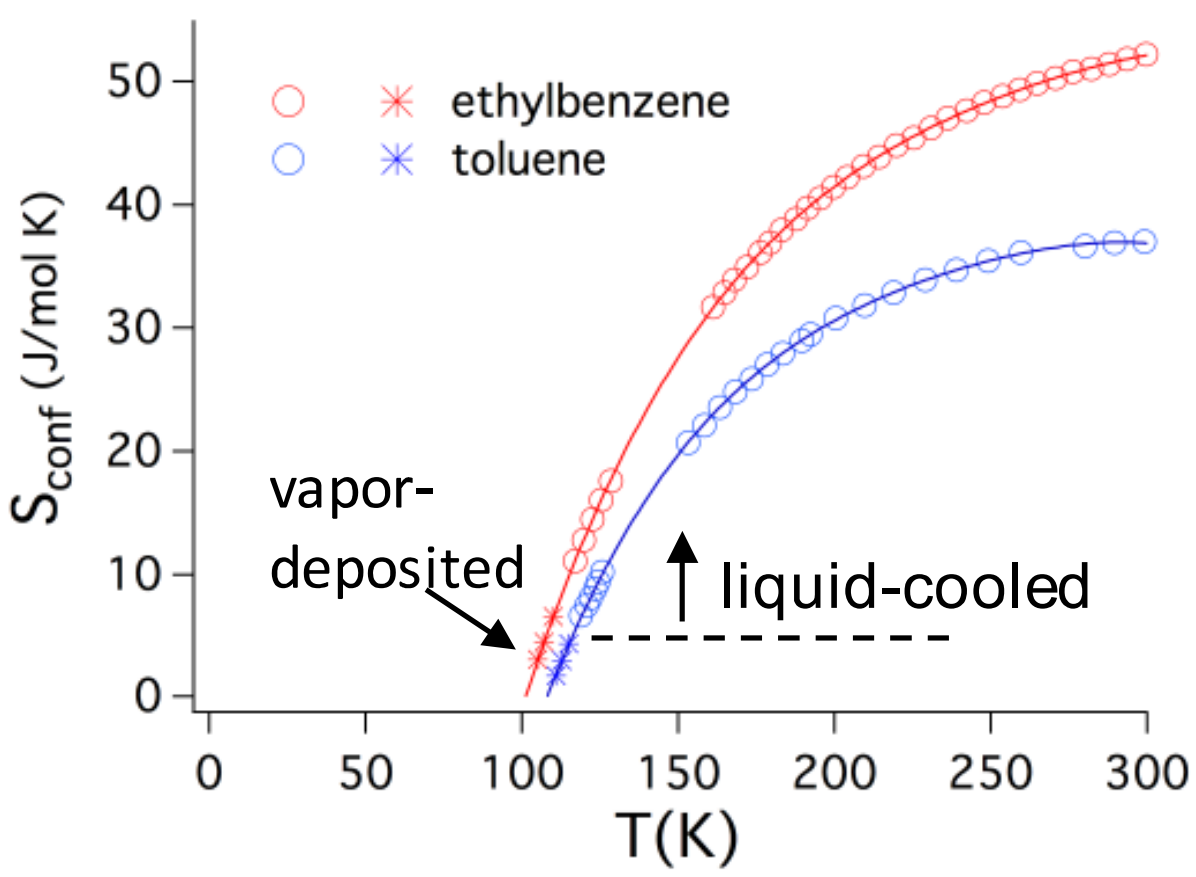

*Figure 7. Configurational entropy ($S_{conf}$) of toluene and ethylbenzene as a function of temperature. If $S_{conf}$ reaches zero, a second order phase transition is expected. Open symbols were obtained from experiments on liquids and supercooled liquids, as reported in reference [36]. Asterisks are calculated from experiments on vapor-deposited glasses of toluene[34, 35] and ethylbenzene[32] in the thermodynamic control regime.*

The combined data sets shown in Figure 7 indicate that the configurational entropies of supercooled ethylbenzene and toluene drop precipitously as $T_K$ is approach from above. Accepting the view that $S_{conf}$ cannot become negative, these data indicate the striking possibility of a second order phase transition at $T_K$. While this conclusion could have been reached without the vapor-deposited data, the addition of these data strengthens the case for a second order phase transition considerably. If somehow a phase transition is to be avoided, the extrapolation shown must fail between $T_K$ and 1.03 $T_K$ for toluene, and between $T_K$ and 1.04 $T_K$ for ethylbenzene. Even if no phase transition occurs, it is clear that vapor-deposition can produce glasses that approach the bottom of the potential energy landscape, as indicated by the small values of the configurational entropy that can be achieved. Two caveats should be mentioned with respect to Figure 7. The calculation of $S_{conf}$ requires assumptions[37] that have a significant influence on the estimate of $T_K$. To put an upper bound on the possible error, we could construct Figure 7 using the excess entropy ($S_{ex} = S – S_{crystal}$); the qualitative features of the figure would remain the same but $T_K$ values would shift down by ~ 10 K, with $S_{ex}$ = 9 J/mol K for the lowest energy vapor-deposited glass. We also note that, for other molecules which yield high stability glasses via vapor deposition, $T_K$ is not so closely approached in the thermodynamic control regime.[16]

The discussion above shows the potential of vapor deposition to provide important new information about states very low in the energy landscape. It is possible that other systems, likely those with higher surface mobility, will allow exploration of states even closer to $T_K$. There has been significant recent progress in understanding the factors that control surface mobility for organic glasses[38] and this work should be used to select systems for future study. Measurements of the enthalpy using a range of deposition rates would be particularly useful. Based upon the work of Rodriguez-Viejo and coworkers, fast scanning nanocalorimetry seems like a particularly promising method for obtaining the enthalpy of vapor-deposited glasses.[39]

The results shown in Figure 7 are qualitatively consistent with recent simulation studies. Ninarello et al. have described a new swap Monte Carlo method that allows the preparation of equilibrium supercooled liquids down to much lower temperatures than previously possible.[40] For a polydisperse mixture of spheres, it has been estimated that these systems are more equilibrated than a liquid-cooled experimental glass (as judged by the number of decades by which dynamics have been slowed). In the simulated systems, $S_{conf}$ decreases to quite low values in a smooth manner analogous to the results shown in Figure 7.[41] Additional simulations provide arguments for a phase transition to the ideal glass at $T_K$ as expected from the mean-field theory of supercooled liquids.[42, 43]

**Diverging time scales?** There has been considerable recent interest in investigating how the relaxation time for a supercooled liquid lengthens as the temperature is lowered. For decades, most experimental data has been fit to the Vogel-Tammann-Fulcher (VTF) equation:

$$\text{Log } \tau/\text{s} = A + B/(T - T_0) \qquad (1)$$

Equation 1 indicates that relaxation times are expected to diverge to infinity at a temperature $T_0$ that can be well above T = 0 K. For some systems, the VTF equation provides a good description of the experimental data over many decades (up to 16 decades)[44] in relaxation times. These results have often been taken to support the idea of a diverging relaxation time at finite temperature. On the other hand, the VTF equation fails to describe data near $T_g$ for a number of systems.[45] Additionally there are alternate fitting functions that adequately fit the available equilibrium relaxation time data but do not indicate divergence at low temperature.[46]

Vapor-deposited glasses have the potential to provide new insight about relaxation times for amorphous systems low in the potential energy landscape. Efforts along these lines are just beginning and conclusions are not yet definitive. As illustrated below, this effort is complicated by the observation that the relaxation times of vapor-deposited glasses can be so long that they cannot be directly measured on a reasonable laboratory time scale.

Figures 8 and 9 show experiments that can be used to estimate the relaxation time of a supercooled liquid well below the conventional $T_g$. For Figure 8, methyl-m-toluate was vapor-deposited onto an interdigitated electrode at various substrate temperatures and dielectric relaxation spectroscopy was used to monitor the stabilities of the as-deposited glasses.[47] For the upper panel, a glass of high kinetic stability was formed at $T_{substrate}$ = 142 K (0.84 $T_g$). In these experiments, no evolution of the sample properties could be detected when the sample was held at $T_{substrate}$. When the temperature was raised 34 K, the sample was observed to transform into the supercooled liquid over a period of roughly 1 hour. This is observed as an increase in the amplitude of the dielectric loss peak; at long times, the response curve is that of the equilibrium supercooled liquid. In the lower panel of Figure 8, the time dependence of the dielectric loss amplitude is plotted for samples deposited over a range of substrate temperatures. We define the isothermal transformation time to be the time required to complete the transformation process as shown in the lower panel. The most stable glass maintained its structure about 5000 times longer than a liquid-cooled glass at the same annealing temperature.

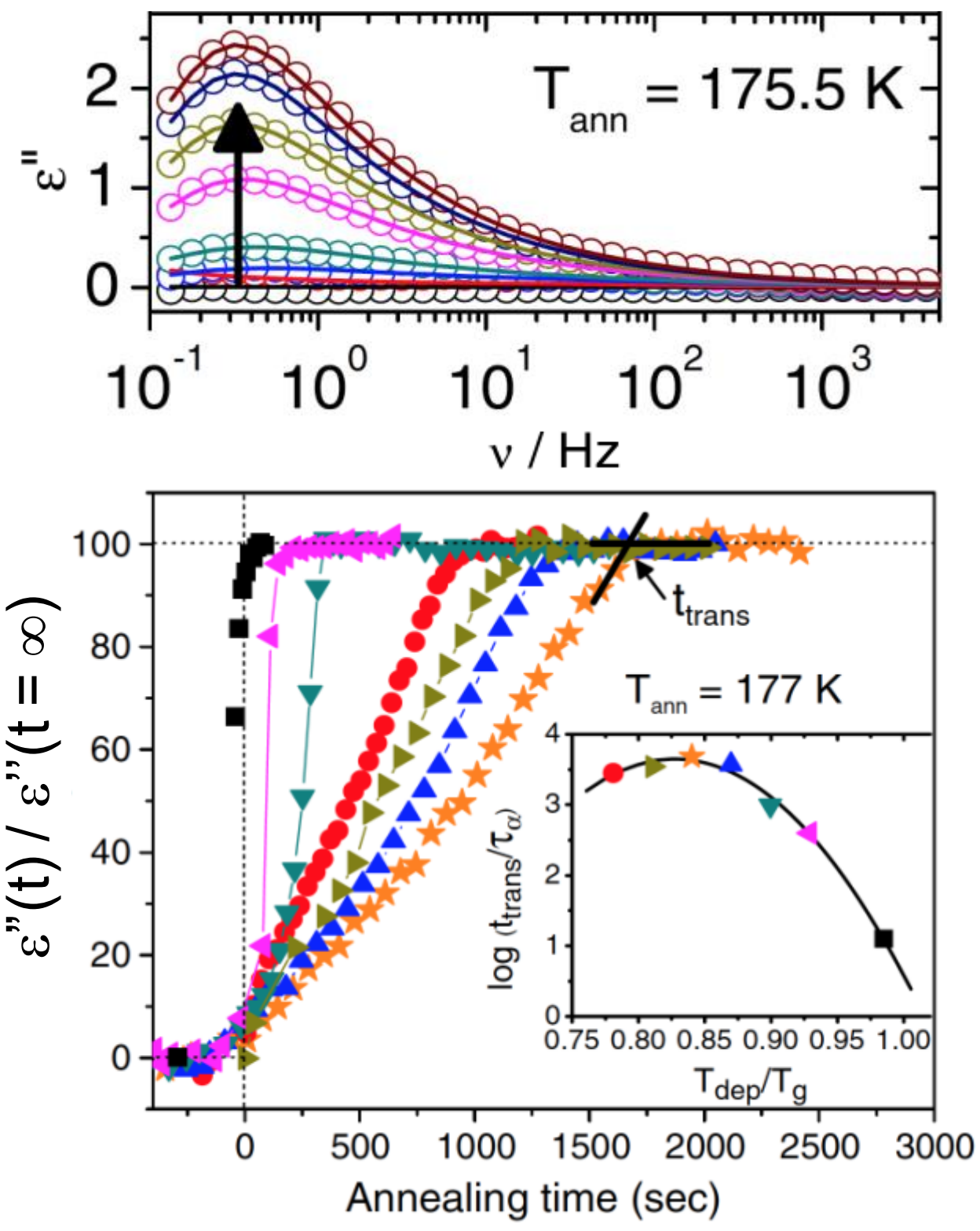

*Figure 8. Dielectric characterization of thick vapor-deposited glasses of methyl-m-toluate. a) Dielectric loss response ($\varepsilon''$) of a sample deposited at $T_{substrate}$ = 142 K during subsequent isothermal annealing at 175.5 K, with the arrow indicating increasing time. b) Evolution of the peak value of $\varepsilon''$ with annealing time for glasses deposited at various $T_{substrate}$. The inset shows the time required to transform glasses deposited at different $T_{substrate}$ into the supercooled liquid, with time normalized to the structural relaxation time of the supercooled liquid at $T_{anneal}$. Reproduced with permission from Phys. Rev. Lett.* ***113****, 045901 (2014). Copyright 2014 American Physical Society.*

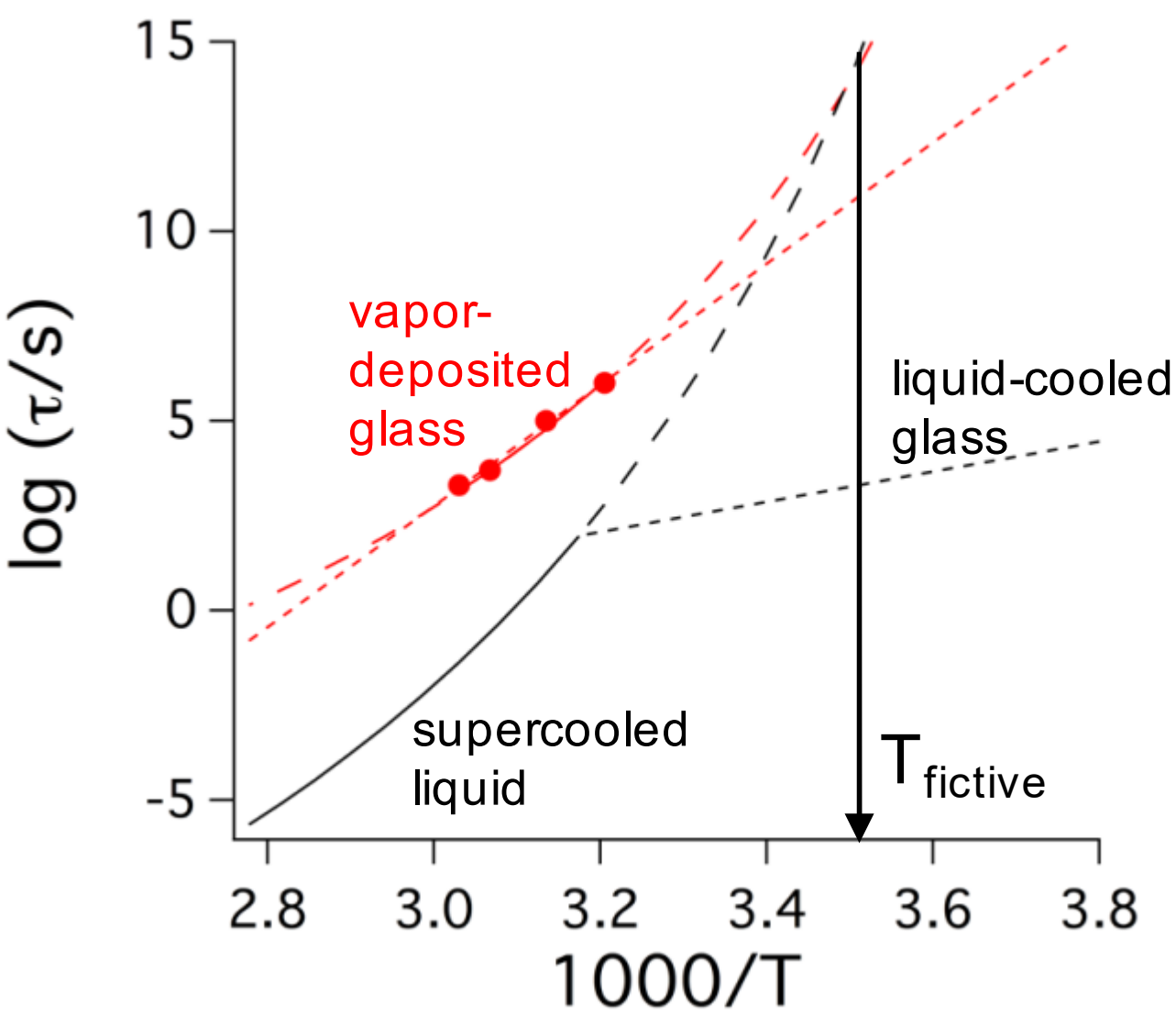

*Figure 9. Use of isothermal transformation times for a vapor-deposited glass of indomethacin to estimate the structural relaxation time of the supercooled liquid at low temperature. Solid red points indicate isothermal transformation times for an indomethacin glass vapor-deposited at $T_{substrate}$ = 265 K from ref [48, 49], with Arrhenius (short dashed) and non-Arrhenius (long dashed) extrapolations to lower temperature; the fictive temperature of this glass as determined by calorimetry[16] is indicated. At $T_{fictive}$, the relaxation time of the vapor-deposited glass is estimated to be at least $10^{11}$ s. For comparison, the solid black line shows a VTF fit to dielectric relaxation data[50] for the supercooled liquid of indomethacin with extrapolation to lower temperature shown as the long-dashed line. The short dashed black line is an estimate of the relaxation time of a glass prepared by cooling the liquid at 10 K/min.*

Isothermal transformation experiments such as the those shown in Figure 8 can be used to estimate the relaxation time of the equilibrium supercooled liquid below the conventional $T_g$.

Figure 9 shows the results of a series of experiments in which multiple samples of a highly stable glass of indomethacin were isothermally transformed into the supercooled liquid at different annealing temperatures (solid red points). At lower annealing temperatures, the transformation time is much larger. In order to make a comparison between the non-equilibrium vapor-deposited glass and the equilibrium supercooled liquid, we make use of the concept of the fictive temperature, i.e., we compare non-equilibrium and equilibrium states with the same enthalpy. The conceptual idea is this: if vapor-deposition produced the equilibrium supercooled liquid at $T_{fictive}$, then the extrapolation of the observed transformation times to $T_{fictive}$ would yield the structural relaxation time of the supercooled liquid.[49] Clearly the required extrapolation is substantial and we show both an Arrhenius extrapolation, and a non-Arrhenius extrapolation that has been shown by Rodriguez-Viejo and coworkers to accurately describe behavior across a wide range of annealing temperatures.[51] Considering these two extrapolations, we estimate the structural relaxation time of the vapor-deposited glass (and the equilibrium supercooled liquid) to be in the range of $10^{11} – 10^{15}$ s ($10^3 – 10^7$ years) at $T_{fictive}$. To put these results into context, Figure 9 also shows the structural relaxation time of the supercooled liquid of indomethacin and its extrapolation to lower temperature according to the VTF equation.[50] The estimated relaxation time of a glass prepared by cooling the liquid at 10 K/min is also shown and has a much weaker temperature dependence. For the vapor-deposited sample, the estimated relaxation time at $T_{fictive}$ exceeds that of the liquid-cooled glass by at least a factor of $10^7$, indicating substantially higher barriers for relaxation. The upper end of the estimated range of times from the experiments is consistent with the VTF extrapolation but clearly this comparison cannot be used as a strong argument in favor of the accuracy of the VTF equation. Pogna et al. have used a completely independent method to estimate the relaxation times of vapor-deposited indomethacin glasses.[52] Using the measured elastic properties, they estimate that the most stable glass has a relaxation time ~ $10^{10}$ s at its fictive temperature. While this value indicates very high activation barriers, it is somewhat lower than the estimates shown in Figure 9.

Estimating the structural relaxation time of an extremely stable glass (or a very deeply supercooled liquid) is clearly a challenging experimental problem. This is an area where new approaches are needed, perhaps utilizing measurement methods that can detect extremely small changes in properties. A recent study of a 20-million-year-old amber glass by Zhao et al.[53] showed the utility of highly sensitive mechanical measurements. It would be very interesting to see these methods applied to vapor-deposited glasses. It would also be useful to test the extrapolation schemes shown in Figure 9 on liquid-cooled glasses that have been aged; some preliminary efforts along these lines have been made with fast-scanning nanocalorimetry.[54]

Computer simulators who generate supercooled liquids at very low temperatures by Monte-Carlo methods also face difficulty in determining extremely long relaxation times.[40] At present, they cannot run molecular dynamics simulations of sufficient duration to be able to directly measure the structural relaxation times. Because the simulations have access to all the particle positions, it may be possible to find robust methods for estimating extremely long relaxation times that might provide guidance for the interpretation of experimental data. Regardless of whether or not the experiments and simulations are consistent with a diverging relaxation time, these efforts provide important information about the energy barriers that need to be surmounted in order for molecules to rearrange in the lower regions of the potential energy landscape.

**Kinetic facilitation.** The idea that an immobile region in a supercooled liquid or a glass can rearrange only if mobility is present in an adjacent region is known as kinetic facilitation.[55-58] This idea provides a useful way to think about how dynamics evolve in a spatially heterogeneous system. Highly stable vapor-deposited glasses exhibit kinetic facilitation to an extent that greatly exceeds anything that occurs in a liquid-cooled glass. When a thin film of a highly stable glass is heated, it transforms via a propagating front of mobility as illustrated in Figure 10.[59, 60] As shown in the upper schematic, the front starts at the free surface and then moves into the glass, with a sharp interface between the not-yet-transformed stable glass (SG) and the supercooled liquid (SCL) that results from the transformation process. The main part of the figure shows the position of the front as a function of time during the transformation for two highly stable glasses of indomethacin that differ only in initial thickness. In both cases, the front emerges from the free surface and propagates at a constant velocity that is independent of sample thickness. For comparison, at this annealing temperature, a liquid-cooled glass would require less than 1 s to transform into the supercooled liquid state; thus the packing achieved by vapor deposition has enhanced kinetic stability by a factor of 10000.

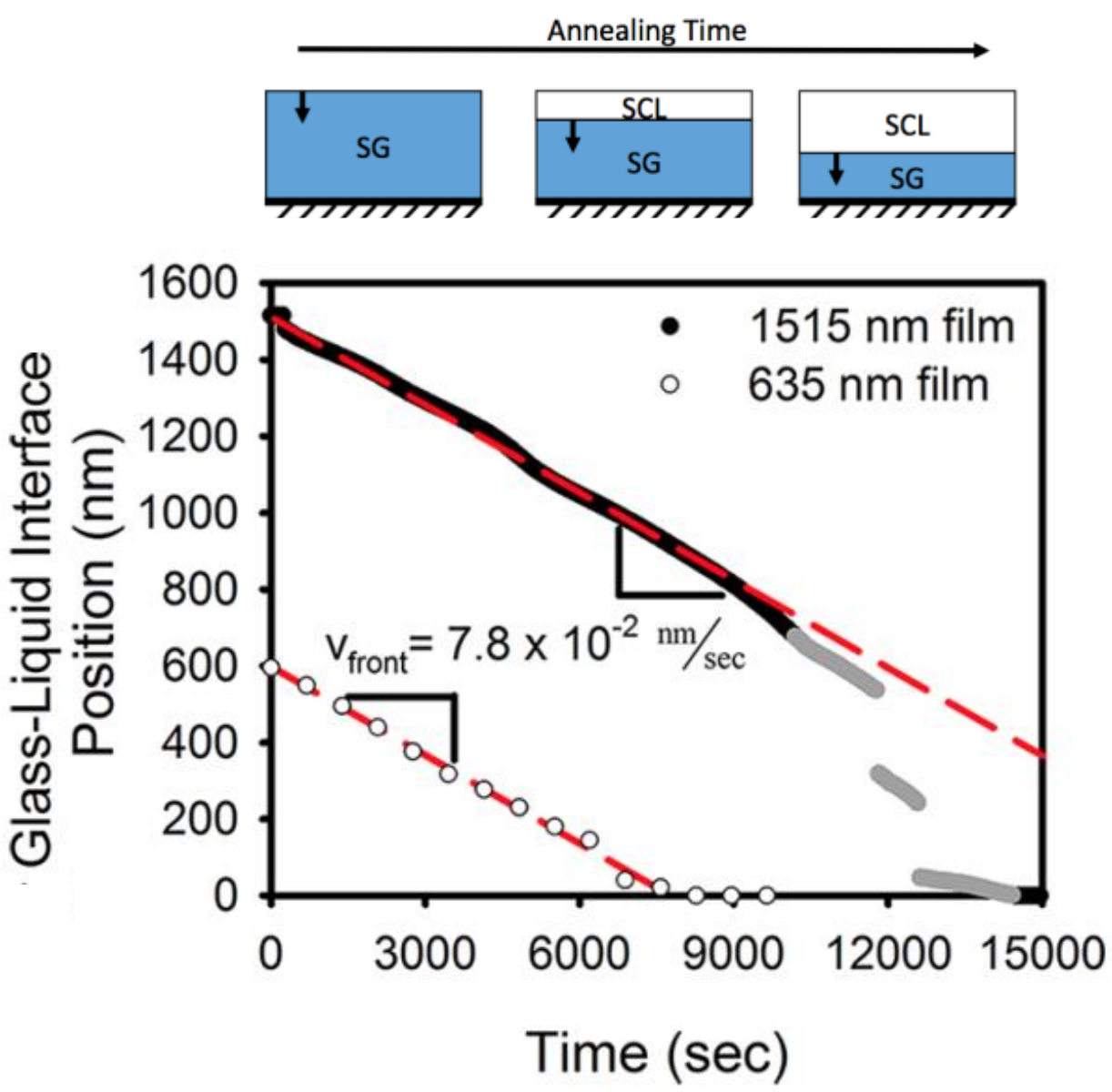

*Figure 10. Transformation of a highly stable glass of indomethacin by a front propagating from the free surface, as detected by spectroscopic ellipsometry.[60] The front propagates at the same velocity in films of two thicknesses. After 10000 s, a bulk mechanism accelerates the transformation process of the thick film. Reproduced with permission from J. Phys. Chem. B **119**, 3875-3882 (2015). Copyright 2015 American Chemical Society.*

We can interpret the propagating front that transforms a stable glass into the supercooled liquid in terms of kinetic facilitation. Given that the free surface of glasses (including stable glasses[61]) can be highly mobile, it is reasonable that the conversion of the stable glass to the supercooled liquid begins at the surface. The supercooled liquid molecules created in this way can then free adjacent molecules from the stable glass. A constant velocity front will emerge as this process happens again and again at the liquid/stable glass interface. The growth front propagation shown in Figure 10 is very similar to the melting of a crystal: in both cases, the transformation to the liquid phase is thermodynamically favored everywhere in the sample, but the kinetic barriers to transformation are much lower at the interface than in the interior. The features of front propagation have been examined theoretically and observed in both atomistic simulations and schematic models of vapor-deposited glasses.[28, 62-64] A front that propagates a long distance from a free surface into a stable glass will eventually be disrupted by sections of the sample that have already been transformed in a bulk-initiated process. For the thicker film in Figure 10 this occurs at a crossover distance of about 1 μm; crossover distances as large as 5 μm have been observed in other systems.[47]

The observation of front propagation in vapor-deposited glasses was initially very surprising because this type of transformation mechanism had not previously been reported for any glass. Some might interpret this to mean that vapor-deposited glasses are fundamentally different materials than liquid-cooled glasses. On the other hand, it may be that the barriers in the bulk of a liquid-cooled glass are so low that a front propagating from the free surface is very quickly disrupted by rearrangements that have already occurred in the bulk material. This second view indicates that propagating fronts might be observed in liquid-cooled glasses in experiments with high time resolution and excellent spatial resolution. A recent fast-scanning (3 $10^4$ K/s) nanocalorimetry study on vapor-deposited glasses of toluene by Rafols-Ribe et al.[39] determined the crossover distances for glasses deposited at different substrate temperatures. For the least stable vapor-deposited glasses, fronts propagated no more than 50 nm. Very recent work by Cubeta et al.[65] reports front propagation for supercooled liquids of toluene and 2-propanol upon heating at a rate of $10^5$ K/s. For toluene, the front was observed to propagate at a much lower velocity and with a higher activation energy for a stable vapor-deposited glass. These recent studies are important because they connect vapor-deposited and liquid-cooled amorphous materials, and seem to indicate that the properties of supercooled liquids and glasses smoothly evolve as stability is increased. From this perspective, the front propagation first observed in vapor-deposited glasses is not a unique feature of materials made by vapor-deposition but rather behavior expected for any amorphous system that is low in the energy landscape. It is also expected that deeply supercooled computer liquids will transform via propagating fronts and this will be a further exciting point of contact between simulation and experiment.

There are some important questions about how the transformation of a stable glass occurs inside a thick sample. One can imagine that regions of poorer-than-average packing (a type of "defect" or "soft spot" in the midst of otherwise efficient packing) would be the first to rearrange upon heating. This would create mobile regions that might initiate a spherical growth front – a growing bubble of the supercooled liquid inside the stable glass (as shown in Figure 4 of reference [66]). How big do these bubbles get? Experiments have been interpreted to indicate growth to more than a micron[66] and a large length scale is consistent with the behavior of some simple models[64, 67] and the RFOT theory.[63] On the other hand, simulations of highly equilibrated supercooled liquids that are subsequently heated do not show evidence of such large length scales.[68] A further problem with the "growing bubbles" picture is that this process should result in a supercooled liquid under pressure. At ambient pressure, the liquid created by transformation is less dense than the stable glass it replaces; unless the stable glass matrix ruptures, this scenario would be expected to yield a bubble full of pressurized liquid. But Figure 8 shows that regions of supercooled liquid grow into a thick stable glass with the relaxation time of the ambient pressure supercooled liquid and hence they cannot be under pressure. This is an area that could benefit from new experiments and simulations.

**Properties of the ideal glass**. The ability to prepare glasses near the bottom of the amorphous part of the potential energy landscape allows us to learn about the properties of very well packed glasses – equivalently, we can (by extrapolation) understand the characteristics expected for the ideal glass. Let us assume for the sake of argument that very slow cooling of the equilibrium supercooled liquid will lead to the ideal glass at $T_K$, as suggested by Figure 7. Figure 5 shows that glasses of indomethacin that are 1.4% more dense than the liquid-cooled glass have already been prepared by vapor deposition; extrapolation[16] to $T_K$ indicates that another 1.4% increase in density is the maximum achievable at pressures near 1 bar. What would be the modulus of such a glass? The most stable vapor-deposited glass of indomethacin has a shear modulus 19% higher than the liquid-cooled glass;[69] by extrapolation to $T_K$, another 19% increase in modulus is expected for the ideal glass.

What molecular motions will be possible in the ideal glass? While the main structural relaxation process will be too slow to be measured, glasses have secondary relaxation processes that occur on shorter time scales.[70, 71] The Johari-Goldstein $\beta$ process in toluene, for example, can be detected by dielectric relaxation and NMR. In Figure 11, the $\beta$ process is compared for liquid-cooled and vapor-deposited glasses of toluene.[72] A stable glass of toluene was vapor-deposited onto an interdigitated electrode. Run 1 shows the dielectric loss at 1000 Hz for the as-deposited glass as obtained upon heating. Above 120 K, the sample begins to transform into the supercooled liquid and the transformation is completed by 130 K. The supercooled liquid was then cooled to low temperature and the dielectric loss of the liquid-cooled glass was measured upon heating (Runs 2 and 3). From solid-state NMR measurements on liquid-cooled glasses of toluene,[73, 74] it is known that the β process involves the reorientation of toluene molecules by an average of 7° on the millisecond time scale; this is a collective process of a group of molecules making subtle adjustments to their packing without being able to escape the cages formed by their nearest neighbors. Figure 11 shows that the β process (the shoulder at about 110 K) is suppressed by a factor of 3.4 in the vapor-deposited glass of toluene. This is consistent with an average molecular reorientation of 3° rather than 7°. This example vividly illustrates the influence of better packing on molecular motion in the glassy state. Moreover, by considering the fictive temperature of the vapor-deposited glass, one can reasonably conclude that the β process would be eliminated in the ideal glass.

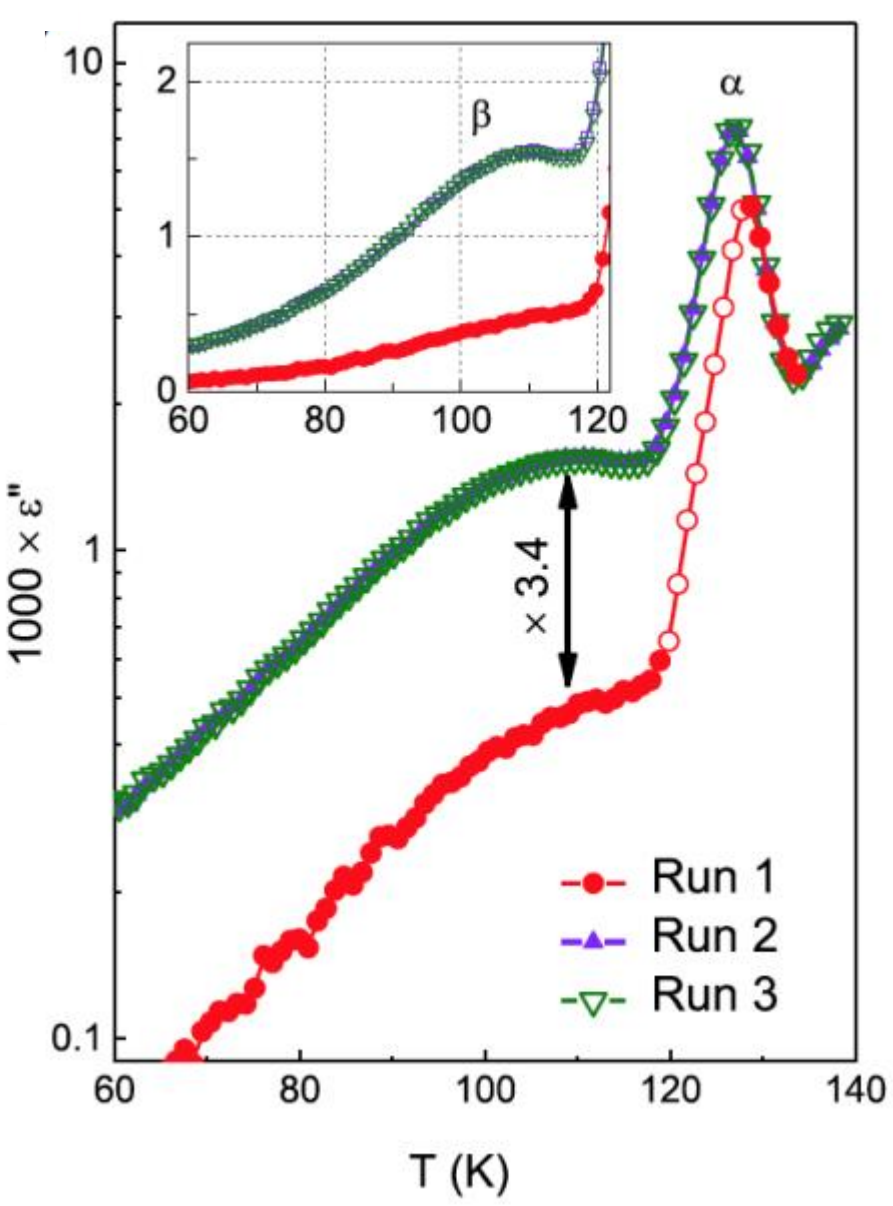

*Figure 11. Suppression of the Johari-Goldstein $\beta$ process for a stable glass of toluene (vapor-deposited at 98 K). The dielectric loss at 1000 Hz is plotted as a function of temperature for the as-deposited glass (Run 1) and the liquid-cooled glass (Runs 2 and 3). The $\beta$ process is the shoulder at 110 K while the main structural relaxation ($\alpha$ process) is the peak at 127 K. The main panel is a logarithmic presentation of the data while a linear plot is shown in the inset. Open red circles are data obtained during the transformation of the stable glass into the supercooled liquid. Reproduced with permission from Phys. Rev. Lett.* ***115****, 185501 (2015). Copyright 2015 American Physical Society.*

While computer simulations may not be able to provide access to the β relaxation in deeply supercooled liquids, preliminary comparisons have been made with other dynamic measures of packing efficiency. The Debye-Waller factor is a measure of the localization of a molecule on a short time scale and provides an indication of the extent to which molecules can explore new positions in the glass without exchanging nearest neighbors. Computer simulations indicate that Debye-Waller factors are suppressed in glasses that are low in the energy landscape, in qualitative agreement with Figure 11.[28]

Recent experiments allow some speculative insights into the properties expected for an ideal glass at low temperatures. At temperatures near 1 K, the heat capacity of a glass is typically much larger than that of the corresponding crystal.[75] This extra heat capacity has been interpreted as due to the relaxation of quantum tunneling two-level systems (TLS).[76, 77] Initially it was thought that the densities of TLS were very similar for all glasses. However, recent evidence from vapor-deposited glasses is consistent with the idea that TLS might be eliminated if amorphous packing could be optimized. Figure 12 shows heat capacity measurements of vapor-deposited indomethacin glasses, in comparison with the liquid-cooled glass and the crystalline material.[78] For the crystal, $C_p/T^3$ is flat at low temperature, consistent with the Debye model. The heat capacity of the conventional liquid-cooled glass is much higher than that of the crystal. Importantly, the vapor-deposited glasses (USG-1 and USG-2) show significantly lower heat capacities than the liquid-cooled glass and follow the Debye model at the lowest accessible temperatures.

A decreased density of TLS could be a general feature of systems approaching the ideal glass state. Related results have been reported for vapor-deposited silicon by Queen et al.[79] For silicon, when $T_{substrate}/T_g$ is increased from 0.36 to 0.74, the density of TLS decreases by nearly a factor of 100; based upon comparison with organic systems, it is anticipated that the packing of the silicon glasses is improved with increased $T_{substrate}/T_g$ but additional experiments here would be very helpful. Theoretical work by Wolynes and coworkers predicts that better equilibrated glasses should have lower TLS levels.[13] However, other factors may influence the comparisons shown in Figure 12. The authors of reference [78] suggest that anisotropic packing may be responsible for the low heat capacity of vapor-deposited glasses. In support of this view, recent experiments on naturally aged amber glasses (with age estimated at 110 million years) show no significant reduction in the low temperature heat capacity,[80] even though the aged sample shows much lower enthalpy than the reference liquid-cooled amber glass. Regardless of its origin, the exciting observation that the density of TLS can be manipulated by vapor deposition seems certain to stimulate further work that will provide new insights into their microscopic origin.

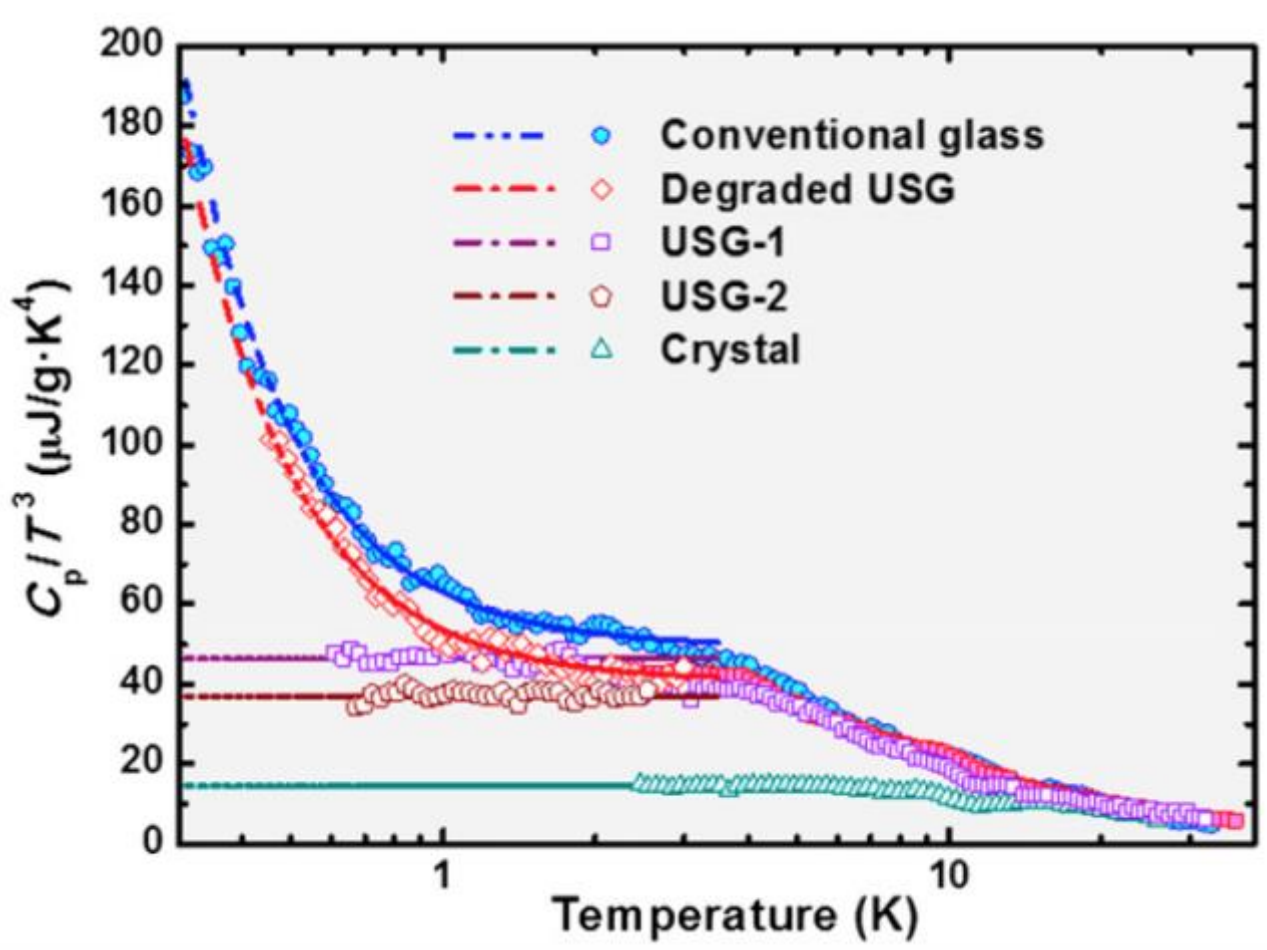

*Figure 12. The heat capacity of indomethacin at low temperature, for the crystal, the conventional liquid-cooled glass, and two vapor-deposited glasses (USG-1 and USG-2).*

*Reproduced with permission from Proceedings of the National Academy of Sciences, 11275-11280 (2014).*

**Other tests of glass stability.** There are many ways to test the stability of a glass. Figures 8-10 show experiments in which temperature is used to test the stability of glass packing; better packing is indicated either by higher temperatures being required to trigger rearrangements, or by a longer timescale for structural relaxation at a given temperature. Progress has been made with other approaches as well. Smith et al. used the permeation of Kr gas into vapor-deposited glasses of toluene and ethylbenzene to probe the onset of molecular motion upon increasing temperature.[81] In good agreement with other work on these systems,[82] they found that glasses deposited onto substrates at ~ 0.92 $T_g$ were most resistant to permeation by the inert gas. Rodriguez-Tinoco et al. investigated whether low enthalpy glasses can better resist crystallization.[83] They studied crystal growth rates on the surface of vapor-deposited celecoxib glasses and found that the glass with the highest thermal stability also exhibited the slowest crystal growth rates (by a factor of 30%). So far, there have been no reports that bulk crystallization can be significantly suppressed by the efficient glass packing arrangements achieved by vapor deposition. Given the importance of controlling crystallization in amorphous pharmaceuticals and organic electronics, this deserves further exploration. Dissolution kinetics might also be influenced by glass stability.

While it is known that crystalline packing can inhibit chemical reactions that occur readily in less ordered environments,[84] only recently has it been demonstrated that glass packing can significantly modulate chemical reactivity. Qiu et al. prepared a highly stable glass of an azobenzene derivative and probed the extent to which photochemical stability can be increased by efficient packing in the amorphous state.[85] They found that the glass with the highest density (prepared by deposition at 0.88 $T_g$) was 50 times more photostable than the liquid-cooled glass and that photostability was highly correlated with glass density. Computer simulations indicate that photostability in this system may result from the inability of the photo-excited azobenzene to reach the *cis* configuration when it is packed into a dense glass.[85] Photostability is a key concern in organic electronics and it will be important to test the generality of these observations, particularly with reactions that have a smaller activation volume than the *trans* to *cis* isomerization of the azobenzene derivative. It would also be of interest to explore whether chemical reactions with atmospheric gases might be suppressed by efficient glass packing. It has already been shown that highly stable glasses of indomethacin absorb 5 times less water vapor than liquid-cooled glasses;[86] lower gas solubility could slow down any chemical reactions with the glass molecules.

A very interesting recent experiment compared the response of vapor-deposited and liquid-cooled glasses of indomethacin to pressure changes.[87] While the vapor-deposited glass had much higher thermal stability at ambient pressure, and both glasses exhibited enhanced thermal stability at higher pressure, the difference in thermal stability between the two glasses disappeared at 300 MPa. The authors established that the effect of pressure was reversible, i.e., upon returning to ambient pressure, the vapor-deposited glass exhibited the same high onset temperature that it showed immediately after deposition. This last observation shows that well-packed glasses can survive density cycling (with increases up to ~4%) without losing the efficient packing conferred by vapor deposition. The similar kinetic stability of the two glasses at high pressure is certainly intriguing and further work exploring the evolution of kinetic barriers under pressure is warranted. Related investigations utilizing nonlinear mechanical experiments, such as nanoindentation, would also be useful.

**Generality of stable glass formation.** In this section, the generality of stable glass formation for organic molecules via physical vapor deposition is considered. Producing stable glasses from materials other than low molecular weight organic molecules is a very important goal and recent progress is summarized.

To date, about 50 organic molecules have been vapor-deposited under conditions where stable glass formation is expected. Roughly 35 of these molecules form highly stable glasses, including both strong and fragile liquids,[47] and molecules with a wide range of shapes[88, 89] and polarities[90]. Many of the properties of these systems are quite similar when compared on a reduced temperature scale. This behavior is illustrated by Figure 13 where the kinetic stabilities of vapor-deposited glasses of 7 organic systems are compared.[91] For many of these systems, similar behavior is observed, with the optimal substrate temperature being 0.8 – 0.9 $T_g$, and the maximum onset temperature being between 1.04 and 1.06 $T_g$; for reference, $T_{onset}/T_g$ is approximately unity for a liquid-cooled glass. For some of the molecules shown in Figure 13, deposition at lower temperatures ($T_{substrate}/T_g < 0.6$) results in the formation of unstable glasses.

Presumably lack of surface mobility at these low temperatures results in "fluffy" structures that readily rearrange upon heating.

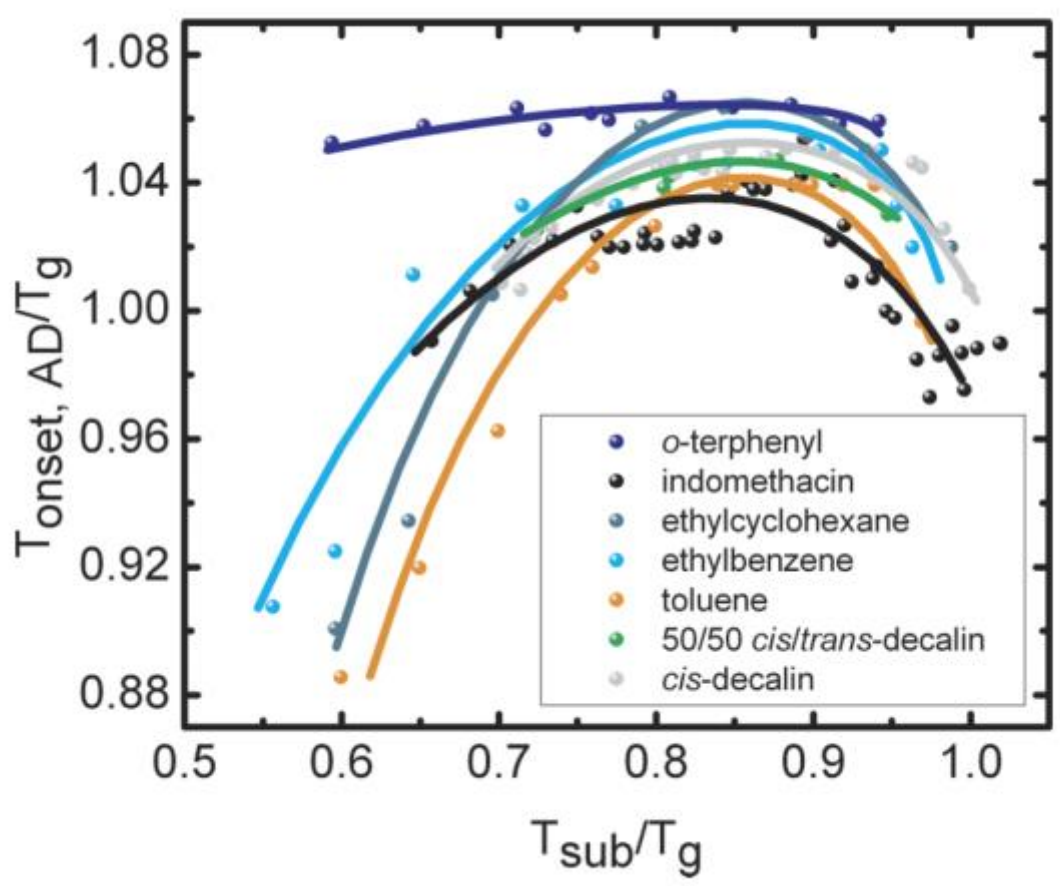

*Figure 13. Kinetic stabilities of vapor-deposited glasses of several organic molecules (and one mixture) as a function of substrate temperature during deposition normalized to $T_g$ for each system. The y-axis shows the onset temperature at which an as-deposited glass begins to transform into the supercooled liquid. $T_{onset}$ values greater than $T_g$ indicate increased kinetic stability relative to the liquid-cooled glass. For most systems, the greatest kinetic stability is exhibited when $T_{substrate}/T_g = 0.8 – 0.9$. All glasses shown were vapor-deposited at 0.15–2 nm/s. Reproduced from J. Chem. Phys.* ***143****, 084511 (2015), with the permission of AIP Publishing.*

When considering the subset of organic molecules that fail to form stable glasses under standard deposition conditions ($T_{substrate}/T_g$ ~ 0.85, deposition rate ~ 0.1 nm/s), there appears to be three different reasons for the failure. Some poor glassformers (i.e., systems that readily nucleate crystals upon cooling the liquid below $T_m$) such as pure *trans*-decalin also form crystals during deposition.[92] Already in 1981, Turnbull recognized that metallic glasses would have mobile surfaces and that as a result nucleation would be possible at the surface during condensation.[93] Organic molecules nucleate crystals much less readily than pure metals and this is presumably the reason why many organic molecules can form glasses when vapor-deposited.

A second failure mode was revealed recently when vapor deposition of a number of hydrogen-bonding systems failed to form glasses with high stability. For example, vapor-deposited glasses of ethanol, propanol, propylene glycol, ethylene glycol, and 2-ethyl-1-hexanol show little or no kinetic stability across the range of substrate temperatures shown in Figure 13.[90] In subsequent experiments, 2-ethyl-1-hexanol glasses were prepared at very low deposition rates and substantially increased kinetic stability was observed, indicating that stable glass formation at the standard deposition rate is thwarted by insufficient surface mobility.[31] This conclusion is supported by a recent study showing that glasses formed from hydrogen-bonding liquids have smaller surface diffusion coefficients than those observed for systems without hydrogen-bonding.[94] This result has been explained as follows: For non-hydrogen-bonding systems, the presence of a free surface means that half of the interactions responsible for slow dynamics are missing, and thus molecules at the surface move much faster.[20] In contrast, hydrogen-bonding systems restructure at the free surface to preserve a large fraction of the hydrogen bonds, and so surface mobility in such systems is not as high.[94]

The most interesting “failure” to form stable glasses resulted in the formation of unusual liquids. In these experiments, Wubbenhorst and coauthors[95, 96] used vapor deposition to prepare glasses of glycerol and higher polyols. While some elements of their observations are consistent with stable glass formation, others are not. More importantly, they reported that vapor deposition of glycerol can form a new liquid state which is characterized by longer relaxation times and a higher dielectric constant. Slow heating to 40 K above $T_g$ succeeded in transforming the new liquid into the standard supercooled liquid of glycerol. Reference [95] also showed that related 4- and 5-carbon polyols (threitol and xylitol) similarly show indications of new liquid states. It was recently reported[97] that the 6-carbon polyol in this sequence (mannitol) exhibits polyamorphism, i.e., the existence of two liquid states connected by a first order phase transition. Further work may show that the systems studied in reference [95] also are polyamorphic. Vapor deposition may have an important role to play in the discovery of new polyamorphic systems. This might be evidenced by a discontinuity in the enthalpy (or density) of vapor-deposited glasses as a function of substrate temperature (in contrast to the behavior shown in Figures 5 and 6).

There have been at least two successful efforts to make interesting polymer glasses by deposition methods. Guo et al. used a laser desorption technique to prepare glasses from polymethylmethacrylate with $M_w$ = 15,000 g/mol.[98] These materials exhibited very high kinetic stability but, in contrast to the low molecular weight systems discussed above, had high enthalpy and low density in comparison with the liquid-cooled glass. While the origin of this combination of properties is still being investigated, these very unusual materials may have important applications. Very recently Yoon et al. prepared a glass of an amorphous fluoropolymer with $M_w$ ~ 100,000 g/mol.[99] In this case, vacuum thermal pyrolysis was used to degrade the starting polymer with repolymerization occurring after deposition. Deposition with $T_{substrate}$ ~ 0.8 $T_g$ prepared materials with remarkably low enthalpy but it is not yet clear to what extent these materials exhibit high kinetic stability.

There have been several efforts to produce highly stable metallic glasses using deposition techniques. Bulk metallic glasses have at least three components and these are the systems being investigated. Yu et al.[100] used a magnetron sputtering approach to deposit glasses with the composition $Zr_{65}Cu_{27.5}Al_{7.5}$. Deposition onto substrates near 0.75 $T_g$ produced glasses with somewhat enhanced kinetic stability ($T_{onset}/T_g$ = 1.016) but surprisingly these materials had higher enthalpy than the liquid-cooled glass. Magagnosc et al. deposited glasses of $Pd_{77.5}Cu_6Si_{16.5}$ using magnetron sputtering and reported that both the modulus and the hardness were maximized for deposition onto a substrate near 0.73 $T_g$.[101] Remarkably, varying $T_{substrate}$ changed the modulus by a factor of 1.7. Recent work by Cao et al. reported high surface diffusion coefficients on the surface of a $Pd_{40}Cu_{30}Ni_{10}P_{20}$ glass,[102] so it seems possible that the mechanism of stable glass formation in metallic systems is also tied to high surface mobility.

Expanding the range of materials from which stable glasses can be prepared is one of the most interesting frontiers for research. A very recent paper reported that vapor deposition of a chalcogenide glass ($Sb_2Se_3$) resulted in enhanced stability against crystallization.[103] There are additional classes of glassy materials (oxides, colloids) for which one could imagine schemes that might be successful in producing high stability materials. For some of these systems, surface mobility during deposition has already been characterized[104] while for other systems surface mobility might need to be artificially enhanced by some means. The key feature of vapor deposition is that it can reduce the time required to reach thermodynamically-favored amorphous states low in the potential energy landscape. There may be quite different approaches that achieve this same goal, e.g., for some chalcogenides, illumination with light of the correct wavelength dramatically lowers the sample enthalpy.[105]

**Structure of vapor-deposited glasses.** Information about the structure of vapor-deposited glasses comes primarily from X-ray scattering and ellipsometry. Most systems that have been studied show indications of anisotropic packing. The notion of an anisotropic glass is troubling for some scientists. Indeed, if one imagines that a glass must inherit its structure from a liquid, then one expects that the isotropic nature of the liquid will be passed to the glassy state upon cooling (see Figure 1). On the other hand, it has been known for decades that cooling a liquid crystalline phase can kinetically trap the structure of the anisotropic liquid into an anisotropic glassy solid.[106] And it is well known that polymer glasses become anisotropic when deformed, either elastically or plastically. It is reasonable to expect that vapor-deposited glasses will often be anisotropic since they are non-equilibrium solids (and hence sensitive to preparation details) and their preparation is anisotropic (deposition from one side). It appears that Hellman and Gyorgy first reported the anisotropy of a vapor-deposited glass, showing in 1992 that vapor-deposited amorphous films of Tb-Fe were magnetically anisotropic.[107, 108] Lin was apparently the first to establish the anisotropy of vapor-deposited organic glasses in 2004, using ellipsometry experiments to show anisotropic optical absorption.[109] Since then, ellipsometry has been used extensively to characterize anisotropy in vapor-deposited glasses.[1, 88, 89, 110-112]

Recent experiments have explored the connection between stability and anisotropy. In particular, synchrotron X-ray scattering techniques have been used to determine the structure of vapor-deposited glasses of organic semiconductors. Figure 14 shows grazing-incidence X-ray experiments on glasses of TPD deposited at various $T_{substrate}$ and also scattering data for a liquid-cooled glass; TPD is a hole transport material used in OLEDs.[113] The highest and lowest $T_{substrate}$ values exhibit clear indications of anisotropic packing while the sample deposited at 300 K has scattering that is nearly isotropic and very similar to the scattering of the liquid-cooled glass. All three vapor-deposited TPD glasses exhibit very high kinetic stability, so anisotropic structure does not seem to be a requirement for high kinetic stability. In support of this interpretation, both highly anisotropic and nearly spherical molecules (e.g., carbon tetrachloride[114]) have been found to form highly stable glasses. Recent simulations reported a strong correlation between stability and potential energy, independent of anisotropy.[115] In recent work by Rodriguez-Tinoco on indomethacin, it was found that highly anisotropic glasses are somewhat less kinetically stable than more isotropic glasses if the comparison is made at the same fictive temperature.[51]

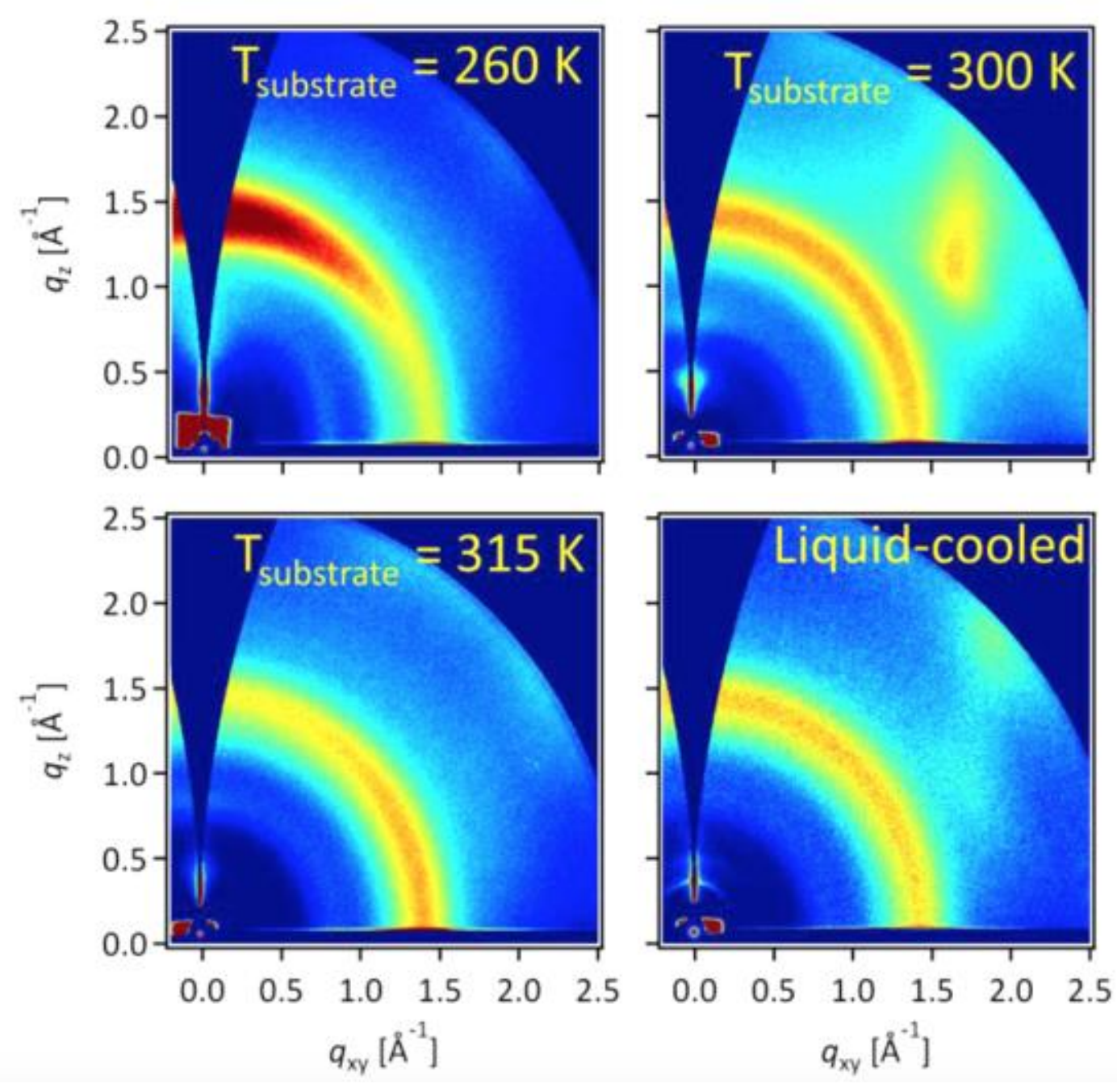

*Figure 14. 2D X-ray scattering patterns obtained from vapor-deposited TPD glasses with a comparison to the liquid-cooled glass. Glasses deposited at 260 K and 315 K show clear indications of anisotropic packing, with a tendency towards "face-on" packing observed at the lower $T_{substrate}$. The glass vapor-deposited at $T_{substrate}$ = 300 K and the liquid-cooled glass exhibit very similar scattering, despite having quite different thermal stabilities. For reference, $T_g$ = 330 K for TPD. The feature in the upper right of the images in the right column is an artifact originating from the silicon substrate. Reproduced with permission from Chem. Mater.* ***27****, 3341-3348 (2015). Copyright 2015 American Chemical Society.*

Does anisotropy contradict the idea that vapor-deposited glasses can be models for equilibrium supercooled liquids and the ideal glass? One expects that the equilibrium supercooled liquid of most organic molecules will remain isotropic down to very low temperatures. While anisotropic vapor-deposited glasses cannot be identical in structure to such a supercooled liquid, it is expected that many properties of the vapor-deposited glass and the equilibrium supercooled liquid will be similar when the fictive temperatures are the same. For example, vapor-deposited glasses of indomethacin shown in Figure 5 are anisotropic below $T_{substrate}$ = 0.97 $T_g$. Nevertheless these glasses have the density, modulus, and enthalpy expected for the equilibrium supercooled liquid down to $T_{substrate}$ = 0.90 $T_g$.[23] It seems unlikely that these mildly anisotropic glasses are much different than the supercooled liquid in this temperature range. Vapor-deposited glasses provide an unprecedented opportunity to learn about states low in the potential energy landscape, even if these states are not identical to the equilibrium supercooled liquid; at present, there is no other laboratory technique for preparing such materials.

Anisotropic glasses are important in the context of organic electronics.[1] For OLEDs, the highest efficiency is achieved if the emitter molecules have their transition dipoles lying in the plane of the device.[116] Photons emitted by out-of-plane transition dipoles are preferentially emitted in the plane of the device and trapped by total internal reflection. In-plane transition dipoles, as opposed to an isotropic distribution, can increase the output of an OLED by a factor of 1.5.[1, 116-118] Increased charge carrier mobility in vapor-deposited semiconductors has been attributed to anisotropic packing;[119, 120] it is reasonable that $\pi$–$\pi$ stacking interactions might be optimized with anisotropic packing. Surface charge measurements indicate that vapor-deposited glasses can exhibit polar order; in such cases the surface potential grows linearly with film thickness up to values that can exceed 100 V.[121-123] This polar order controls the polarization at the interface between two semiconductors which in turn influences charge injection in an operating device.[124]

Considerable progress has been made recently in understanding what controls the anisotropy of organic glasses prepared by vapor deposition. While a full exploration of these issues is beyond the scope of this article, the surface equilibration mechanism discussed above can rationalize the anisotropy of vapor-deposited glasses if one accounts for the anisotropy of the equilibrium supercooled liquid at the free surface.[125] This idea originated from computer simulations of the deposition process[88, 89, 126] and is illustrated schematically in Figure 15. The left-side illustrates the structure of the equilibrium liquid of a rod-shaped molecule. Color-coding delineates regions of different molecular orientation in the equilibrium liquid, with purple indicating isotropic packing in the interior of the liquid. During deposition, molecules near the surface strive to equilibrate towards the structure of the equilibrium interface. At low $T_{substrate}$, only the top layer succeeds and this leads to glass in which horizontal orientation is trapped. At a higher $T_{substrate}$, the top two layers manage to equilibrate during deposition; further deposition then leads to glass in which vertical orientation is trapped.

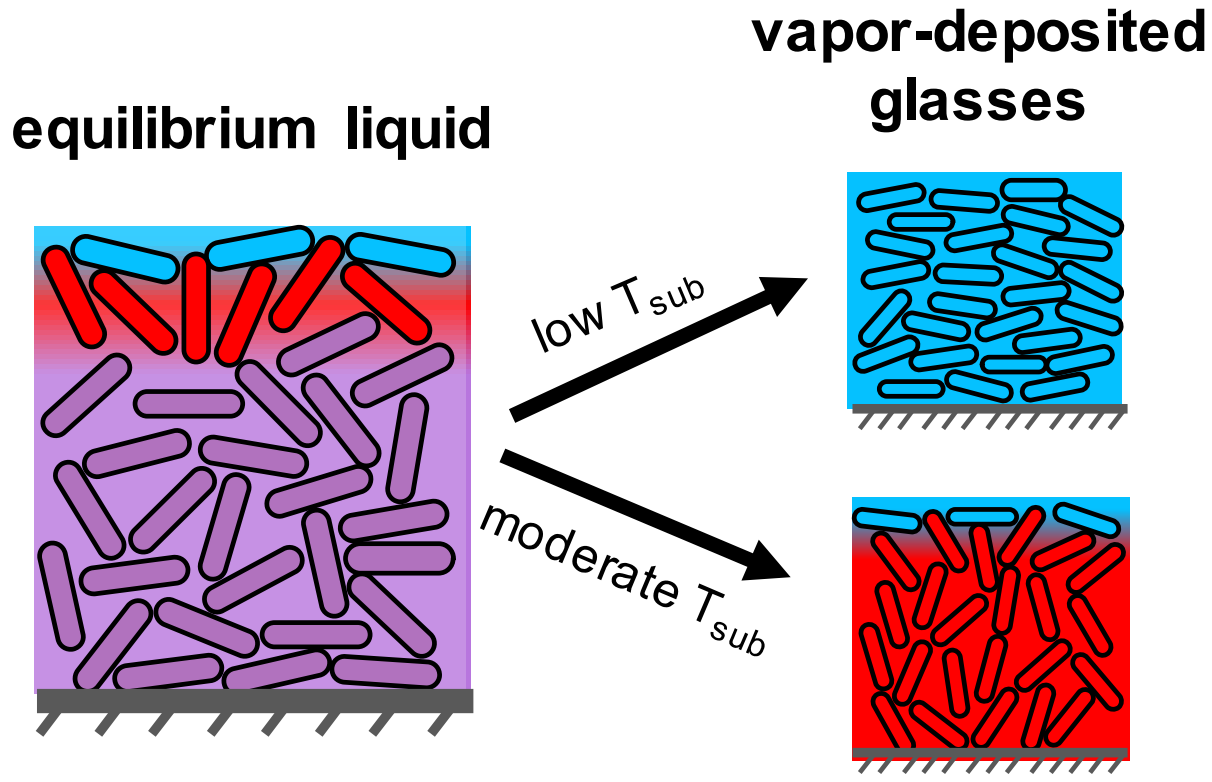

*Figure 15. Schematic showing the origin of anisotropic molecular packing in vapor-deposited glasses, based upon a detailed analysis of computer simulations.*[88, 89, 126] *The lozenge shapes illustrate the orientation of a rod-like molecule such as TPD. The free surface of the equilibrium liquid is anisotropic. The substrate temperature determines the depth to which structure at the surface can equilibrate during deposition. The lowest portion of the equilibrated structure is trapped by further deposition.*

There are exciting new efforts to further manipulate the anisotropy of vapor-deposited organic glasses. Deposition of liquid crystalline mesogens can result in very highly ordered glasses and the substrate temperature can provide considerable control over the orientation of the structures prepared.[127] There have been several studies examining polar order in vapor-deposited glasses for both organic semiconductors and model compounds.[121-123, 128] And there is the possibility of active control over anisotropy during deposition through external fields. Further work along these lines will produce new types of anisotropic organic solids in which molecular packing can be manipulated over a wide range of structures. These materials have considerable potential for use in organic electronics, optoelectronics, and other fields. It seems likely that the surface equilibration mechanism will provide a helpful framework for interpreting results from all of these areas.

**Concluding remarks.** This is an exciting time for research in supercooled liquids and glasses. For the first time, vapor deposition allows laboratory preparation of glasses that are low in the potential energy landscape. New advances in simulation provide access to somewhat analogous low energy computer glasses. Other recent developments in simulation and theory are providing a refined understanding of how the entropy crisis is resolved. Already experiments have uncovered previously unsuspected features of low energy glasses including transformation via propagating mobility fronts. Understanding the transformation mechanism of bulk samples of low energy glasses is an important area for future research. An important goal is the development of better experimental methods for calculating the configurational entropies of supercooled liquids as this limits our ability to quantify the position of any glass relative to the bottom of the energy landscape.

Vapor deposition is one example of a method that provides efficient access to the lower regions of the energy landscape. By assembling the glass from the surface and making use of enhanced surface mobility, it is possible to bypass the high barriers that must be overcome during slow cooling of a liquid or isothermal annealing of a glass. In a somewhat analogous manner, Monte Carlo computer simulations speed equilibration by utilizing moves that the real system cannot access. It would be a major step forward if additional experimental approaches could be found that allow the preparation of low energy glasses. There is no reason why these new approaches need to utilize vapor deposition or assembly from the surface.

There are significant opportunities for advances in the structural characterization of vapor-deposited glasses. There are important x-ray and neutron experiments that have not yet been utilized including polarized resonant x-ray scattering (for investigating the spatial extent of orientation correlations) and the determination of the radial distribution function. It would be helpful to use advanced electron microscopy methods that provide higher order structural information characterizing local packing arrangements. Computer simulations should be able to integrate all of this information and produce reasonable real space models for packing in low energy vapor-deposited glasses.

**Acknowledgements**. MDE gratefully acknowledges support from the National Science Foundation (CHE-1564663 and DMR-1720415) and the US Department of Energy, Office of Basic Energy Sciences, Division of Materials Sciences and Engineering, Award DE-SC0002161. MDE thanks Lian Yu, Steve Swallen, Juan de Pablo, Christoph Schick, Ranko Richert, Bob McMahon, Sushil Satija, George Fytas, and Michael Chabinyc for collaboration on vapor-deposition projects along with former and current students. MDE thanks Diane Walters, Madeleine Beasley, and Kushal Bagchi for assistance with figures.